\documentclass[useAMS,usenatbib]{mn2e}

\usepackage{graphicx}
\usepackage{times}
\usepackage{natbib}

\begin{document}

\title[Thermal and non-thermal traces of AGN feedback] {Thermal and non-thermal traces of AGN feedback:  results from cosmological AMR simulations}
\author[F. Vazza, M. Br\"{u}ggen, C.Gheller]{F. Vazza$^{1,2,3}$\thanks{E-mail: franco.vazza@hs.uni-hamburg.de}, M. Br\"{u}ggen$^{1,2}$, C. Gheller$^{4}$\\
$^{1}$ Jacobs University Bremen, Campus Ring 1, 28759, Bremen, Germany, \\
$^{2}$ Hamburger Sternwarte, Gojenbergsweg 112, 21029 Hamburg, Germany\\
$^{3}$INAF/Istituto di Radioastronomia, via Gobetti 101, I-40129 Bologna,
Italy\\
$^{4}$  Swiss National Supercomputing centre, Via Cantonale, CH-6928 Lugano, Switzerland}

\date{Received / Accepted}
\maketitle
\begin{abstract}
We investigate the observable effects of feedback from Active Galactic
Nuclei (AGN) on non-thermal components of the intracluster medium
(ICM). We have modelled feedback from AGN in cosmological simulations
with the adaptive mesh refinement code {\small ENZO}, investigating three types of
feedback that are sometimes called quasar, jet and radio
mode.  Using a small set of galaxy clusters simulated at high
resolution, we model the injection and evolution of Cosmic Rays, as
well as their effects on the thermal plasma.  By comparing, both, the
profiles of thermal gas to observed profiles from the {\small ACCEPT}
sample, and the secondary $\gamma$-ray emission to the available upper
limits from {\small FERMI}, we discuss how the combined analysis of
these two observables can constrain the energetics and mechanisms of
feedback models in clusters.  Those modes of AGN feedback that provide
a good match to X-ray observations, yield a $\gamma$-ray luminosity
resulting from secondary cosmic rays that is about 10 times below the
available upper limits from {\small FERMI}. Moreover, we investigate
the injection of turbulent motions into the ICM from AGN, and the
detectability of these motions via the analysis of line broadening of
the Fe XXIII line.  In the near future, deeper
observations/upper-limits of non-thermal emissions from galaxy
clusters will yield stringent constraints on the energetics and modes
of AGN feedback, even at early cosmic epochs.

\end{abstract}

\label{firstpage} 


\section{Introduction}
\label{sec:intro}

Radiative cooling of gas in galaxy clusters is so efficient that most
of the hot gas phase in their core ought to be removed on a time scale
smaller than the lifetime of the system, producing an inward motion of
the cooling gas, a ''cooling flow''
\citep[e.g.][]{1984Natur.310..733F}. However, dramatic cooling flows
are not observed in real clusters, and additional {\it
non-gravitational} heating mechanisms are likely to keep gas on a
higher adiabat
\citep[e.g.][]{1991ApJ...383..104K,1999Natur.397..135P,2000MNRAS.315..689L}.
It is largely agreed that active galactic nuclei (AGN) are a
viable source of energy available for the
self-regulation of galaxy clusters. Observations show that the energy
associated with AGN in clusters is in most cases sufficient to balance
radiative losses in the ICM \citep[e.g.][and references
therein]{2007ARA&A..45..117M}. However, it is less clear how the
energy is released from the compact ($\ll$kpc) region surrounding the
central super-massive BH, to the $\sim 10-100$ kpc cooling radius.
Very similar problems are encountered in elliptical galaxies
\citep[e.g.][]{1987ApJ...320...32S,1997ApJ...487L.105C,2000ApJ...535..650B,2012ASSL..378...83C}.
For clusters, an important issue is whether most of the energy input
from AGN (or galactic winds) to the ICM has occurred much before the
assembly of the clusters via "pre-heating"
\citep[e.g.][]{2001ApJ...555..597B,2001ApJ...546...63T,2001ApJ...553..103B,mcc2004},
or at a low redshift within already formed clusters
\citep[e.g.][]{1995MNRAS.276..663B,2001ApJ...554..261C,2002Natur.418..301B}. The
first mechanism requires a lower energy budget, with total energies
$\leq 10^{62} \rm erg$ \citep[][]{mcc2008}, while the second
possibility requires energies in excess of $\sim 10^{63}-10^{64} \rm
erg$ \citep[][]{2011ApJ...738..155M}.  However, there is ample
evidence for strong AGN outflows in cool core clusters
\citep[e.g.][]{2007ARA&A..45..117M,2012NJPh...14e5023M}.

Theoretical work suggests that the
real evolution of heating in clusters might be a combination of both
\citep[][]{2006ApJ...643..120B,mcc2008,va11entropy,2012MNRAS.420.2662D,short12}.

 
Quasar-induced outflows at high-redshift, possibly following mergers of gas rich galaxies, have been observed \citep[e.g.][]{2006ApJ...650..693N,2010ApJ...709..611D,2008MNRAS.389...34B}.
At lower redshift, the mechanical work done by X-ray cavities on the surrounding ICM may represent another viable mechanism for heating and mixing the ICM \citep[e.g.][]{2001ApJ...557..546D,2007ARA&A..45..117M,2012NJPh...14e5023M}.
Additional mechanisms can modify the energy requirements of AGN feedback, by providing complementary heating/mixing on various scales. 
The heating from Alfv\'{e}n waves in cosmic rays (CR) enriched bubbles, and heating from Coulomb losses of CRs and the surrounding thermal ICM is an additional interesting topic of research 
\citep[][]{1991ApJ...377..392L,2008MNRAS.387.1403S,2004A&A...413..441C,guo08,ma11,fuj11}. 

Thermal conduction can somewhat reduce the energy budget that central AGN have to provide in order to stem cooling flows \citep[e.g.][]{1986ApJ...306L...1B,1988ApJ...326..639B,2001ApJ...562L.129N}.
Magneto-rotational instabilities and heat-flux driven instabilities in the weak and anisotropic cluster magnetic
field have been proposed to reduce the cooling of gas \citep[e.g.][]{2008ApJ...673..758Q,mcc10}. Finally, also major mergers
have been suggested as a viable mechanisms to reduce the cooling
catastrophe, even if the real efficiency of this mechanism is controversial \citep[][]{2006MNRAS.373..881P,2008ApJ...675.1125B}. 

\bigskip
Cosmological simulations with radiative
 cooling, star formation and galactic winds are unable to
 reproduce the observed profiles of gas temperature, metallicity and entropy in the
 ICM \citep[e.g.][for a
 recent review]{2012arXiv1205.5556K}. 
 The most powerful feedback from AGN has been studied with cosmological simulations in the recent past.  Several groups successfully
 implemented a treatment of thermal AGN feedback in cosmological
 {\small GADGET} simulations
 \citep[e.g.][]{2004MNRAS.355..995D,2006MNRAS.366..397S,2007MNRAS.380..877S,mcc2010,2010MNRAS.401.1670F}. In the aforementioned papers, the growth of BHs at the centres of
 galaxies is followed using sink-particles, and the energy release
 from each BH follows from the Bondi-Hoyle accretion rate ("quasar"
 mode).  The energy emitted by the AGN is released by heating up SPH
 particles surrounding the surrounding ICM. Similar methods have been
 implemented in AMR grid simulations by \citet{2007MNRAS.376.1547C}
 and \citet{teyssier11} in the {\small RAMSES} code.  The creation of
 ''bubbles'' inflated by AGN during their ''radio mode''
 was simulated by \citet{2004MNRAS.355..995D} and
 \citet{2006MNRAS.366..397S} in {\small GADGET}. In this case, the
 energy released by the black hole (BH) is deposited within pairs of
 bubbles in the form of thermal energy or cosmic ray energy, and
 exerts mechanical work $P dV$on the surrounding ICM while buoyantly
 rising in the cluster atmosphere.  Other models of mechanical
 feedback from AGN have also been implemented, by assigning a ''wind''
 drift velocity to gas particles surrounding the BH, with velocities
 in the range $\sim 10^3-10^4$ km/s
 \citep{2008MNRAS.387.1431D,2010MNRAS.401.1670F}.
 \citet{dubois10,dubois11} implemented a scheme to follow bipolar
 kinetic outflows ("jet" mode) from simulated AGN in {\small RAMSES},
 monitoring the growth of BHs using the same setup of
 \citet{teyssier11}.  In the framework of galaxy formation studies,
 run-time models of radiative and kinetic feedback from AGN in galaxy
 simulations have been developed by
 \citet{2011ApJ...738...54K} in {\small ENZO} simulations, and by
 \citet{2006MNRAS.373.1265O} and \citet{2011MNRAS.417.2676G} in
 {\small GADGET} simulations.

To date, however, little attention has been paid to the amount of
non-thermal energy deposited in the ICM by the various feedback
mechanisms. This is an important aspect, since non-thermal emission
from galaxy clusters can offer a complementary way of testing and
falsifying feedback models.  Once accelerated, CR hadrons can
accumulate in galaxy clusters \citep{bbp97} and produce a non-thermal
component that could be detected by $\gamma$-ray observations
\citep[e.g.][] {ack10}.  Secondary particles are continuously injected
into the ICM via proton--proton collisions, possibly leading to
detectable synchrotron radiation \citep[e.g][]{bl99, de00}. The
combined analysis of radio observations and $\gamma$-upper limits,
however, presently suggests that most of the observed large-scale
radio emission in clusters cannot be due to secondary electrons, based
on theoretical estimates of the required total energy in CRs
\citep[][]{br07,donn10,2009A&A...508..599B}, and to the required
values of the magnetic field in the ICM, in conflict with observations
\citep[][]{2009A&A...507..661B,bo10,2010MNRAS.401...47D,bonafede11,jp11}.
 
Therefore, the detection or lack of non-thermal emission from cluster
centres may inform us about the energy budget of non-thermal
particles, and of the history and modality of previous heating
episodes in the ICM.  In addition, CR particles may have an important dynamical effect on the ICM
\citep[e.g.][]{ro11, bl11}, and  they can also affect the evolution of
X-ray cavities powered by AGN jets
\citep[e.g.][]{mb07,2008MNRAS.387.1403S,guo08,ma11}.

\bigskip
In this work we study the observable non-thermal features related to
AGN feedback models in a cosmological framework. A few single-object
simulations have been used to investigate the role of CR feedback in
stopping cooling flows \citep[][]{guo08,fuj11,fuj12}. However, the
acceleration (and re-acceleration) of CRs at shocks triggered by 
AGN, as well as at merger and accretion shocks have been neglected in
previous works.

To our knowledge, the present study is the first in which such
detailed CR physics (e.g. particle acceleration, reduced
thermalization at the sub-shock, pressure feedback of CRs, effective
adiabatic index of the baryon gas) as well as variety of AGN feedback
models (quasar, jet and bubble modes) have been applied to cosmological simulations.
\bigskip

\section{Numerical methods}
\label{sec:methods}

We have produced cosmological cluster simulations with the adaptive mesh refinement code {\small ENZO}. On the basis of the public 1.5 version of {\small ENZO} we have implemented our methods to model the evolution and feedback of CR particles injected at shock waves \citep[][]{scienzo}, as well as our (simplified) implementations of energy release from AGN.

{\small ENZO} is a grid and adaptive mesh refinement (AMR) code using the Piecewise Parabolic Method (PPM) to solve the equations of hydrodynamics, 
originally written by \citet{br95} and developed by the Laboratory for Computational
 Astrophysics at the University of California in San Diego \citep{no07,co11}
{\footnote {http://lca.ucsd.edu}.  

The detailed description of our modules for CR-physics and AGN feedback are presented in Sec.2.1-2.1. 
In all runs in this paper, we adopted 
radiative cooling for a fully ionized H-He plasma with a constant metallicity of $Z=0.3 Z_{\odot}$, and a cooling function with a cut-off at $T=10^{4}$ K \citep[][]{1987ApJ...320...32S}, as in the public version of {\small ENZO}, while the re-ionization background due to the UV radiation from
early stars and AGN  is modelled by keeping a gas temperature floor ($\sim 2 \cdot 10^4$ K) in the redshift range $4 \leq z \leq 7$ \citep[as in][]{va10kp}.

We did not include star formation and feedback through winds or supernovae in these runs, 
thereby reducing the complexity and memory usage of the code. Hydrodynamical simulations suggest that while 
supernovae are important to reproduce the observed metal
distribution of the ICM, they do not have a significant impact on the
thermal history of the ICM on large scales \citep[e.g.][]{short12}.

For the simulations presented here, we assumed a ``concordance'' $\Lambda$CDM cosmology with
$\Omega_0 = 1.0$, $\Omega_{B} = 0.0441$, $\Omega_{DM} =
0.2139$, $\Omega_{\Lambda} = 0.742$, Hubble parameter $h = 0.72$ and
a normalization for the primordial density power
spectrum $\sigma_{8} = 0.8$.

\subsection{Cosmic ray-physics}
\label{subsec:cr}

The basic methods to model the injection, advection and pressure feedback of CRs in our {\small ENZO} runs are explained in detail in \citet{scienzo}. We assume that CRs are injected at shocks with an acceleration efficiency, $\eta(M)$, that only depends on the Mach number, $M$, which is given by diffusive shock acceleration  
\citep[e.g.][]{be78, bo78, dv81, ebj95, kj90, md01,kj07,2012arXiv1206.1360C}. 
New CR energy is {\it injected} in the system by multiplying the energy flux through each shocked cell by the 
time step and the cell surface (at each AMR-level):

\begin{equation}
 E_{\rm cr}= \eta (M) \cdot \frac{\rho_{\rm u} v_{s}^{3}}{2} \cdot \frac{\Delta t_{\rm l}}{\Delta x_{\rm l}},
\label{eq:secondary}
\end{equation}
where $\rho_{\rm u}$ is the pre-shock density, $v_{\rm s}$  is the shock velocity, $\Delta t_{\rm l}$ and $\Delta x_{\rm l}$ are the time step and
the spatial resolution at the AMR level-l, respectively.  To 
ensure energy conservation the thermal
energy in the post-shock region is {\it reduced} proportionally at run-time. The dynamics of the mixture of gas and CRs within each cell follows the total effective pressure of the two, 
$P_{\rm eff}=P_{\rm g}+P_{\rm cr}=\rho [(\gamma-1)e_{\rm g}+(\gamma_{\rm cr}-1)e_{\rm cr}]$, where $\gamma=5/3$, $\gamma_{\rm cr}=4/3$, $e_{\rm g}$ is the gas energy density and $e_{\rm cr}$ is the CR-energy density. 
The dynamical feedback of CR pressure is treated in the Riemann solver by
updating the gas matter fluxes in the in 1--D sweeps along the coordinate axes and using the effective
gamma factor ($\gamma_{\rm eff}=\frac{(\gamma P_{\rm g}+\gamma_{\rm cr} P_{\rm cr})}{P_{\rm g}+P_{\rm cr}}$), for the computation of the local sound speed in cells.

As in \citet{scienzo}, we use the relativistic value of $\gamma_{\rm cr}=4/3$ everywhere, which corresponds to the flattest possible momentum spectrum
of CRs, through $\gamma_{\rm cr}=q/3$ (with $f(p) \propto p^{-q}$ and $q=4$ for $\gamma_{\rm cr}=4/3$). Fixing the value of $\gamma_{\rm cr}$ within the simulated volume is an unvoidable assumption of the two-fluid model adopted here. However, once the CR-energy density is specified, the CR-pressure depends only weakly on the spectral shape of $f(p)$ and on the cut-off of the spectrum \citep[e.g.][]{1993ApJ...402..560J,ju08}.

\bigskip

To model the injection of CRs at each shocked cell, we developed a run-time shock finder based on
pressure jumps, similar to \citet{ry03}. At each time step, we flag cells
with a negative 3--D divergence, $\nabla \cdot {\vec v} < 0$, and with concordant local gradients of temperature and
entropy, $\nabla S \cdot \nabla T>0$ \citep[e.g.][]{ry03}. The local Mach number is computed by inverting standard
Rankine-Hugoniot jump conditions for gas pressure.  This method can run in either fixed grid resolution or adaptive mesh refinement mode 
in {\small ENZO} runs, and compares very well with the shock finding methods we develop elsewhere \citep[][]{va09shocks}.
Compared to our first paper on this subject \citep[][]{scienzo}, we
add the treatment of several additional physical processes that
are listed below.

\subsubsection{Shock re-acceleration}

We model at run-time the shock re-acceleration of CRs, by shocks running over a medium already enriched
of CR energy by previous injections. This can be particularly relevant in the case of shocks caused by
AGN feedback, where at late redshift the ICM is already enriched with
CRs ($P_{\rm cr}/P_{\rm g} \sim 10^{-2}-10^{-1}$ at $z \leq 0.5$).
According to the results in \citet{kj07} the presence of CRs in the pre-shock region mimics an increased injection efficiency of CR-energy in the post-shock.
This dynamic effect can be treated by using a different analytical function for $\eta(M)$,
dependent on the ratio $P_{\rm cr}/P_{\rm g}$ in the pre-shock. In our case, we calculate $\eta(M)$ via a linear
interpolation between the extreme cases of $E_{\rm cr}/E_{\rm g}=0$ (in which case the efficiency $\eta(M)$ is identical to the one of \citealt{scienzo}) and $P_{\rm cr}/P_{\rm g}=0.3$ \citep[both taken from][]{kj07}. Figure \ref{fig:eta} shows the acceleration
efficiency, based on the interpolation of \citet{kj07}, as a function

of the pressure ratio between CRs and thermal gas in the pre-shock region. The post-shock thermalization is then reduced at run-time accordingly, as in \citet{scienzo}. Based on our tests in \citet{scienzo}, the run-time modelling of shock re-acceleration of CRs does not have a dramatic impact on the final distribution
of CR-energy, that is on average increased by $\sim 10$ percent  inside the virial radius.  This is because the pressure ratio between CRs and thermal
gas in always tiny inside clusters (Sec. 3.2), and therefore the impact of re-accelerated CRs on the final budget of CR-energy of the ICM is small.

\begin{figure}
\includegraphics[width=0.45\textwidth]{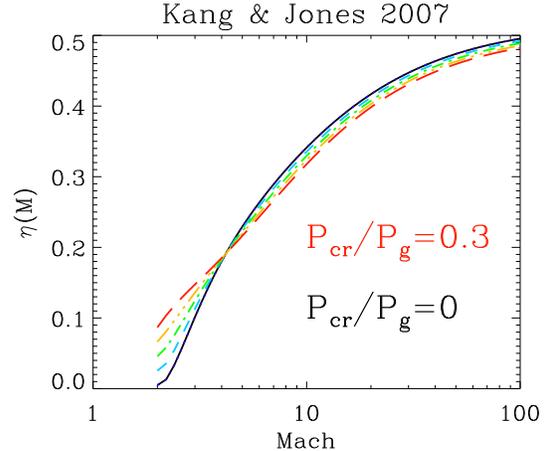}
\caption{Acceleration efficiency of CRs as a function of Mach number for different pre-existing ratios of
$P_{\rm cr}/P_{\rm g}$ in the pre-shock region. From black to red, the different lines show the acceleration efficiency for $P_{\rm cr}/P_{\rm g}=0.0$, $=0.06$, $=0.12$, $0.18$, $=0.24$ and $=0.3$.}
\label{fig:eta}
\end{figure}

\subsubsection{Hadronic and Coulomb losses}
\label{subsubsec:hadronic}

Cosmic rays can lose energy via
binary interactions with thermal particles of the ICM. This channel
of energy exchange between thermal and relativistic particles
in the ICM is important for the high gas 
density ($\rho/(\mu m_{\rm p})>10^{-2} \rm cm^{-3}$) of cool
cores. Relativistic protons transfer energy to the thermal gas via Coulomb collisions with the ionized gas.
They can also interact hadronically with the ambient thermal gas and produce mainly $\pi^+$, $\pi^-$ and $\pi^0$, provided their kinetic energy exceeds the threshold of 282
MeV for the reaction. The neutral pions decay after a
mean lifetime of $\approx 9 \cdot 10^{-17}$ s into $\gamma$-rays.
To estimate the total energy transfer rate between CRs and thermal
gas in both mechanisms, we need to determine
the CR energy spectrum. Since this information is not readily available in the two-fluid model, we must assume an
approximate steady-state spectrum for the CR energy distribution.
We fixed a spectral index of $\alpha=2.5$ for the particle energy{\footnote {This choice of $\alpha$ corresponds to a Mach number of $M=3$ at the particle injection. On average, this represents the typical injection spectra of particles accelerated at powerful merger shocks in cosmological simulations, which are dominant sources of thermalization and CR injection within clusters \citep[e.g.][and references therein]{va11comparison}.}} and computed the total Coulomb and hadronic loss rates as a function of the ICM density and of $E_{\rm cr}$ for
each cell, as in \citet{guo08}:

\begin{equation}
\Gamma_{\rm coll}=-\zeta_{\rm c} \frac{n_{\rm e}}{\rm cm^{-3}} \frac{E_{\rm cr}}{\rm erg \cdot cm ^{-3}} \rm erg \cdot s^{-1} \cdot cm^{-3} 
\label{eq:gamma_coll}
\end{equation}

where $n_{\rm e} \approx n$ is the electron number density,
and $\zeta_{\rm c}= 7.51 \cdot 10^{-16}$ is the coefficient for all collisional energy loss terms. In hadronic
collisions, only $\sim 1/6$ of the inelastic energy goes into secondary electrons
\citep[][]{2004A&A...413..441C,guo08}. The energy-dominating region of CR electrons
($\gamma \sim 10^{2}$) will heat the ICM through Coulomb interactions, plasma oscillations and excitation of Alfv\'{e}n waves
\citep[e.g.][]{guo08}. Therefore we can 
assume that these secondary electrons lose most of their energy
through thermalization and thus heat the ICM. Similar to Eq.\ref{eq:gamma_coll}, the heating rate of the ICM through Coulomb
and hadronic collisions can be computed as:

\begin{equation}
\Gamma_{\rm heat}=\xi_{\rm c} \frac{n_{\rm e}}{\rm cm^{-3}} \frac{E_{\rm cr}}{\rm ergs \cdot cm^{-3}} \rm erg \cdot s^{-1} \cdot cm^{-3}, 
\label{eq:gamma_heat}
\end{equation}

where $\xi_{\rm c}=2.63 \cdot 10^{-16}$ \citep[][]{guo08}.
In our simulations with radiative cooling and AGN feedback, the rate
of energy loss due to these collisions is extremely small, typically $\sim 10^{-3}-10^{-4}$ of $E_{\rm cr}$ or $e_{\rm g}$ during the time step at each AMR level. This allows us to use a simple first-order integration to compute the energy losses of CRs (and the corresponding gas heating rate) at run-time.

In our previous work \citep[][]{scienzo}, we did not include hadronic and Coulomb losses. In general, modelling this process at run-time decreases
the CR energy by a factor $\sim 10$ within cluster cores, while yielding identical results for the remaining cluster volume compared to runs that neglect losses.

\begin{figure*}
\includegraphics[width=0.95\textwidth]{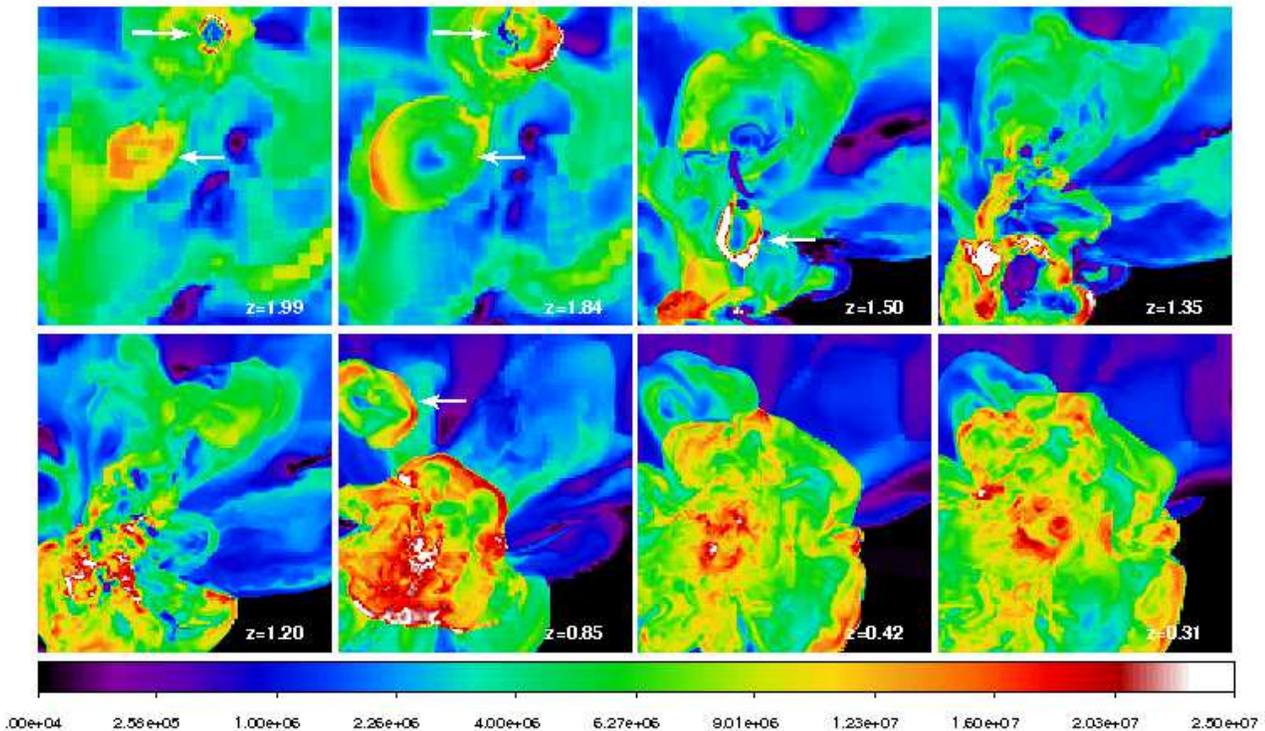}
\caption{Time sequence of gas temperature (in units of [K] in the colour coding) for a slice of $5 \times 5 $ comoving Mpc/h and width $25 ~\rm kpc/h$ for run H5 with "quasar" feedback. The arrows point to the locations of recent "quasar" events in the volume. The "coarse" resolution of the first snapshots is due to the fact that we turn on AMR based on velocity jumps at $z=2$.}
 \label{fig:temp_evol_agn}
\end{figure*}

\begin{figure*}
\includegraphics[width=0.95\textwidth]{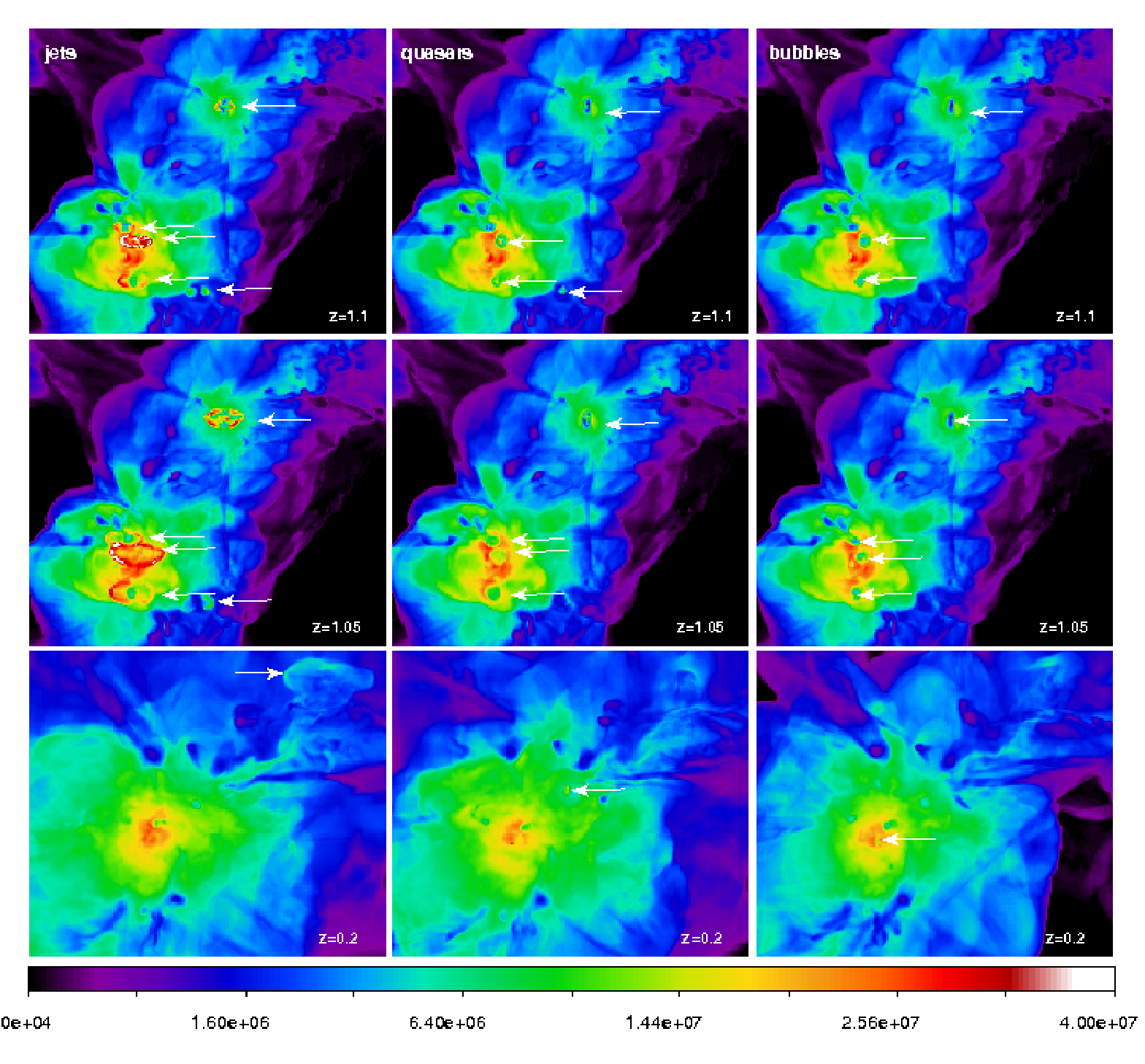}
\caption{Time sequence of mass-weighted average gas temperature for a comoving
 volume of $\sim (5 ~\rm Mpc/h)^3$ in the formation region of cluster H5.
 We show the results of the three re-simulations employing AGN feedback,
 always with an energy per event of $E_{\rm AGN}=10^{59} \rm erg$.
Horizontal arrows suggest the location
of interesting episodes of AGN feedback within the volume.}
\label{fig:temp_cfr}
\end{figure*}

\subsection{Models of feedback from AGN}
\label{subsec:agn}

In \citet{va11entropy} we implemented a simple model of AGN feedback in {\small ENZO}, via
injection of thermal energy at the opposite sides of the cooling
region of galaxy clusters.
Here, we explore more complex recipes of energy feedback 
between the cold gas and the surrounding ICM, allowing also for a direct input of CR energy from AGN.

At each time step, the identification of a suitable location of the central super-massive BH is based on the simple measure of local gas over-density. First, we flag cells hosting a gas density exceeding a given threshold, $n \geq n_{\rm BH}$, and then we select as active ''AGN-cells'' only the maxima within cubic regions of size $\approx 1$ Mpc/h. Based on more detailed modelling of BH growth using sink-particles \citep[][]{2007MNRAS.380..877S,teyssier11,2012MNRAS.422.3081M},  we tuned the threshold value to excite AGN feedback to $n_{\rm min} \approx 10^{-2}  \rm cm^{-3}$. 
The use of this threshold is motivated by the fact that in our fiducial
setup (peak resolution of $25  \rm kpc/h$ per cell) the mass enclosed in a cell with $n=10^{-1} - 10^{-2} \rm cm^{-3}$ is $\approx 10^{10}-10^{11} \rm M_{\odot}$. This is the typical gas
mass surrounding BHs of $M_{\rm BH} \sim 10^{8}- 5 \cdot 10^{9} \rm M_{\odot}$, which are commonly hosted inside the masses of galaxy clusters and groups \citep[e.g.][]{2007MNRAS.380..877S,teyssier11,2012MNRAS.422.3081M}{\footnote {The choice of relying only on the gas density as a proxy may trigger also feedback from cold filaments connecting galaxies, where AGN feedback of the type considered here is unlikely. However, our
simulations have sufficient high resolution only in galaxy clusters. This
seems to exclude any spurious release of feedback energy from cold filaments,
as indeed we find. To fully avoid this possibility, one would have to resort
to more complex models (e.g. sink particles).}}.

In the
following, we will refer to the cells exceeding this density threshold
and powering energy feedback as ''AGN-cells''.

\bigskip

We have implemented three modes of AGN feedback: a ''quasar'' mode (i.e. thermal output of energy from AGN, Sec.\ref{subsubsec:quasar}), a ''jet'' mode (i.e. kinetic energy output from bipolar jets around AGN, Sec.\ref{subsubsec:jets}) and
a ''radio'' mode (i.e. creation of buoyant bubbles in pressure equilibrium with the ICM, Sec.\ref{subsubsec:radio}).

Once the feedback mode is specified, the only parameters that must be set are: a) the initial redshift for the
start of AGN feedback ($z_{\rm AGN}$) and b) the energy release of each single AGN-event, $E_{\rm AGN}$. In the case of "quasar" and "jet" modes this directly measures the energy we provide for each burst of either thermal or kinetic energy, while for the "bubble" feedback this represents the estimated total energy released to the ICM by the creation of bubbles with internal pressure $P_{\rm bb}$ and volume $V_{\rm bb}$, $E_{\rm AGN} \approx 3 P_{\rm bb} V_{\rm bb}/2$ \citep[e.g.][]{2008ApJ...686..927S}.

Even if jets and radio bubbles are associated with the same type of AGN feedback \citep[e.g.][]{2007ARA&A..45..117M,2012NJPh...14e5023M}, in our study they are considered as alternative scenarios. This allows us to distinguish the CR-effects of buoyancy and impulsive
kinetic feedback in a clearer way.

In a preliminary set of tests (see the Appendix) we explored various recipes for the 
implementation of feedback
modes in one reference cluster,  before varying the efficiencies in the whole set
of clusters. 

Here, however, we will discuss only the "fiducial" subset of parameters for which each implementation of feedback modes showed
the best performance. 
While the range of spatial resolution achieved in our runs is probably not sufficient to study specific small-scale
features associated with each AGN mode (e.g. the morphology of jets or rising bubbles), our resolution
and physical setup are suitable for studying the large-scale features of CRs in the ICM.

\subsubsection{Thermal feedback from AGN}
\label{subsubsec:quasar}

A significant fraction of the energy emitted from AGN can thermally couple to the surrounding gas. One can define efficiencies such that the energy added in time $\Delta t$ is
\begin{equation} 
\Delta E_{\rm AGN,g}=\epsilon_{\rm r} \epsilon_{\rm f} {\dot{M}}_{\rm
BH} c^2 \Delta t,
\end{equation} 
where $\epsilon_{\rm r} \sim 0.1$ is the bolometric radiative efficiency  for a Schwarzschild BH \citep[][]{1973A&A....24..337S}, $\dot{M_{\rm BH}}$, and $\epsilon_{\rm f}$ is the coupling efficiency with the thermal gas, which
is usually assumed to be in the range $\epsilon_{\rm f} \approx 0.05-0.15$
in order to fit the observed $M_{\rm BH}$ vs $\sigma_{\rm v}$ relation (e.g. Di Matteo et al.2005; Booth \& Schaye 2009). 
This thermal coupling between the AGN and the surrounding ICM is 
usually called "quasar" feedback, which may be a mechanism for pre-heating of the
ICM \citep[][]{2005ApJ...619...60L,mcc2004,mcc2008,2009MNRAS.400..100S,lapi10}.
In addition, 
quasar-induced outflows at high-redshift, possibly following mergers of gas rich galaxies, have been observed in many cases
\citep[e.g.][]{2006ApJ...650..693N, 2008MNRAS.389...34B, 2010ApJ...709..611D}.
Similar to \citet{mcc2010} and \citet{teyssier11}, when we detect cells with
$n \geq n_{\rm min}$, we release the thermal energy, $E_{\rm AGN}$, adding to the total and internal gas energy inside
cells at the highest available AMR level.

This implementation of quasar feedback is only an approximation of the true physical processes at play, i.e. the launching of strong winds due to the radiation pressure of photons from the accretion disc. This is unavoidable, given that our best resolution is orders of magnitude larger than the theoretical accretion disc region, and also given the difficulty of modelling the radiative transfer of photons from the accretion region.

Our choice of a minimum gas density $n_{\rm min}=10^{-2} \rm cm^{-3}$ selects the typical environment of massive BHs within
clusters and groups. Furthermore, we assume an energy output similar to most theoretical models, without actually measuring the 
accretion power of BHs at run-time. This is different from simulations where the mass growth of BHs is modelled using sink particles, which enables an accurate reconstruction of the BH matter accretion rate at run-time. This is not possible in our case. Our approach is only a first step to include the
effects of AGN feedback in our
version of the code, and it allows us a fast and efficient study of AGN modes.





\bigskip

\subsubsection{Kinetic feedback from AGN}
\label{subsubsec:jets}

The innermost ICM can be affected 
by the injection of kinetic energy through bipolar jets originating from the AGN \citep[e.g.][]{1995MNRAS.276..663B,2012ASSL..378...83C}. 
This energy can be thermalized
by impacting on the ICM after a short distance from the cluster centre ($10-100$ kpc, \citealt{2009MNRAS.395.2317P}). 
\citet{gaspari11a,gaspari11b} and \citet{dubois10,dubois11} recently simulated the mechanical coupling
between purely kinetic jets from the AGN and the surrounding ICM
in the cooling region.

In this model each of the two jets is initialized as a pure input of kinetic energy density $E_{\rm AGN} = 1/2 \rho v_{\rm jet}^2$, 
with  velocity, $v_{\rm jet}$, pointing radially outwards from the cluster centre. 
Even if in our version of {\small ENZO} the launching direction of the jets is set by a random selection of coordinate axes, in this work
we keep the jet axis fixed. In this way, velocity effects related to the direction of jet launching can easily be detected (Sec.\ref{subsec:turbo}).
Every time a jet is generated, we modify
the gas velocity, the total and internal energy at the highest available AMR level in a pair of cells on opposite sides of the gas density peak.
The width  and the initial extension of the jet are set by the maximum resolution, which is $\sim 10$ times larger than the best resolution available in "single-object" runs \citep[][]{gaspari11a}. At our resolution, the equivalent opening angle of each jet
is $\sim 30-40$ degrees with respect to the cluster centre.
This is an unavoidable drawback of the fact that in such
cosmological runs achieving a much larger resolution is 
computationally very expensive.
 However, for the dynamical feedback of jets the most important quantity is the injected kinetic energy, which is similar to simulations at higher resolution, once the different mass load of the jets and launching 
 velocity are considered.
Also the initial velocity of our jets (which depends on the
typical density  of AGN-cells, through the total kinetic energy released in the AGN-burst)  is  $\sim 500-1000 ~\rm km/s$, about one order of magnitude
lower than the  typical velocity of jets in single-objects simulations at much higher resolution  \citep{gaspari11a,gaspari11b}.

\subsubsection{Bubble feedback from AGN}
\label{subsubsec:radio}

The creation of buoyant bubbles in the ICM inflated by jets requires a spatial resolution ($\sim 0.1 - 1$ kpc) beyond what can be achieved in our cosmological runs. 
For this reason, we created already formed evacuated
bubbles around cells hosting an AGN, similar to \citet{2008ApJ...686..927S}.}  The bubbles are created in (approximate) pressure
equilibrium with the surrounding ICM, by decreasing the gas
density inside the cells by $\rho'=\delta_{\rm bb} \rho $, and correspondingly increasing the thermal energy in order to conserve the original gas pressure. During run-time, this modification is performed over a single time-step of the simulation, at the highest available AMR level.
This obviously leads to a loss of gas mass in these cluster runs. 
The simulations show that on average this loss amounts to $\sim 5-10$ percent of the gas mass by the end of the simulation.

The initial under-density of the bubbles is set by energy conservation inside the bubbles, after that $E_{\rm AGN}$ is added to the gas and CR energy within the same volume.
For the range of $E_{\rm AGN}$ and $n_{\rm min}$ adopted in this work, the under-density inside the bubbles is typically $0.1-0.01$.  The bubbles are generated as a pair of {\it cubic} blocks of $2^{3}$ cells, at the distance of 2 cells from the cluster centre. This is a very crude approximation and is expected to lead to expedient 
numerically mixing of the thermal energy.


In principle, we can also allow for the presence of CRs
injected by the AGN, parametrized by $\phi_{\rm cr}=E_{\rm cr,AGN}/E_{\rm AGN}$ of the total injected energy, similar to
\citet{2008MNRAS.387.1403S}. In this case the energy of the bubbles is 
conserved also by modifying the effective adiabatic index of the mixture of gas+CRs \citep[as in][]{scienzo}. 
 However, in the main 
body of this paper we only refer to purely {\it thermal} bubbles.
We present a few results with gas+CRs inflated bubbles in
the Appendix (the results, however, are not significantly different). 


\begin{table}
\label{tab:tab1}
\caption{List of the physical models adopted in our runs. Column 1: identification name.
C2: cooling. C3: energy per event. C4: details on feedback mode (with approximate values of heating temperature, jet velocity and initial bubble under-density).}
\centering \tabcolsep 5pt 
\begin{tabular}{c|c|c|c}
  ID & cooling & $E_{\rm AGN}$ [erg] & feedback mode \\  \hline
   cooling  & yes & 0 &  no \\
   quasars & yes &  $10^{59}$ & thermal, $T_{\rm AGN} \sim 5 \cdot 10^{7} K$  \\
   quasars2 & yes & $10^{60}$ & thermal, $T_{\rm AGN}=5 \cdot 10^{8} K$ \\
   jets & yes &  $10^{59}$ & kinetic, $v_{\rm jet} \sim 750 ~\rm km/s$ \\
   bubbles & yes & $10^{59}$ & buoyant, $\delta_{\rm bbl} \sim 0.05$  \\
 \end{tabular}
\end{table}

\begin{table}
\label{tab:tab2}
\caption{Main parameters of the clusters simulated in this work. Column 1: identification name. C2: total (gas+DM) mass at $z=0$. C3: virial radius ($R_{\rm v}$). C4: dynamical state at $z=0$. The identification name of each cluster is choosen 
to be consistent with other works of ours \citep[][]{va10kp,scienzo}. 
The cluster parameters are referred to the simple non-radiative simulation
of each object.}
\centering \tabcolsep 5pt 
\begin{tabular}{c|c|c|c}
  ID & $M_{\rm tot} [10^{14} \rm M_{\odot}]$ & $R_{\rm v} [\rm Mpc]$ & dynamical state\\  \hline
   E1  & 11.2 & 2.67 & post-merger\\
   E5A & 8.2 & 2.39 & merging\\
   H5 &  2.4 & 1.70 & post-merger \\
   H10 & 1.2 & 1.20  & relaxed \\
 \end{tabular}
\end{table}

\section{Results}

The results discussed in the main part of this paper have been produced considering the following physical mechanisms:
a) pure cooling; b) cooling and thermal feedback by AGN from $z \leq 4$, with a fixed energy release per event of $E_{\rm AGN}=10^{59} ~\rm erg$ (or also $10^{60}~\rm erg$ in the case of E1 and E5A) ; c) cooling and kinetic energy feedback by AGN from $z \leq 4$, with a fixed energy release per event of $E_{\rm AGN}=10^{59} ~\rm erg$; d) cooling and energy feedback from buoyant bubbles injected by AGN from $z \leq 4$, with an
approximate energy per vent of $E_{\rm AGN} \sim 10^{59} ~\rm erg$.
With the above implementations, we re-simulated four clusters in 
the masses range $10^{14} \rm M_{\odot} \leq M \leq 10^{15} \rm M_{\odot}$.
 The list of the most important parameters of
all ``fiducial'' models investigated in the article is given in Tab. 1.
The first model (pure cooling) obviously produces strong cooling
flows in all systems. In this work,
this model is therefore regarded only as a standard reference to assess the role
of each feedback model.

In order to achieve the largest possible dynamical range inside the volume where each cluster forms, the four clusters were re-simulated at high spatial and DM mass resolution starting from
parent simulations at lower resolution and adding nested initial conditions
of increasing spatial and mass resolution \citep[e.g.][]{abel98}. Two levels of nested initial conditions were placed in cubic regions centered on the cluster centers. 
The box at the first level had the size 
of $\approx 95 ~\rm Mpc/h$ (with $m_{\rm dm} \approx 5.4 \cdot 10^{9} ~\rm M_{\odot}/h$ 
and constant spatial resolution of $\Delta_{1} \approx 425 ~\rm kpc/h$). The second box had a size
of $\approx 47.5 ~\rm Mpc/h$ (with $m_{\rm dm} \sim 6.7 \cdot 10^{8}~\rm M_{\odot}/h$ and constant spatial resolution of $\Delta_{1} \approx 212 ~\rm kpc/h$). For every cluster run, we identified cubic regions
with the size of $\sim 6 R_{\rm v}$ (where
$R_{\rm v}$ is the virial
radius of clusters at $z=0$, calculated in lower resolution runs), 
and allowed for 3 additional levels of mesh refinement, 
achieving a peak spatial resolution of $\Delta \approx 25~\rm kpc/h$
(in the following, we will refer to this sub-volume as to the ``AMR region'').
From $z=30$ (the initial redshift of the simulation) to $z=2$, mesh refinement is  triggered by gas or DM over-density. From $z=2$ an additional refinement 
criterion based on 1--D velocity
jumps \citep[][]{va09turbo} is switched on. This
second AMR criterion is designed to capture shocks and intense turbulent motions in the  ICM out to the clusters outskirts. Shocks and turbulence are analyzed in the next Sections (Sec.\ref{subsec:shocks}-\ref{subsec:turbo}). In 
addition, our test with {\small ENZO} have shown that this composite AMR 
criterion better capture physical mixing motions in the ICM, and reduces the
amount of numerical mixing, with important consequences in the amount of cold
gas within the cluster volume \citep{va11entropy}.
However, this method is computationally much more
expensive with respect to the gas/DM overdensity criteria alone, and we could
afford to run it only in the ``fiducial'' AGN models (Tab. 1).
 
In the Appendix we show the effects of different implementations and energies budget in AGN feedback in the smallest of these systems,using the standard AMR criterion which works on gas/DM overdensity. In the main paper we will study the 
observable thermal and non-thermal features of the 
"fiducial" implementation of each feedback mode in our version
of {\small ENZO}.  The parameters of the small cluster sample resimulated with the different recipes of AGN feedback is given in Tab. 2.

The evolution of gas temperature between $z=2$ and $z=0$ in cluster H5 (quasar-mode) is shown in the panels of Fig.\ref{fig:temp_evol_agn}. Blast waves triggered by AGN feedback (highlighted by arrows) in the thermal mode drive powerful shocks through the intra cluster medium, adding to the pattern of merger and accretion shock waves. We observe that in general only $2-3$ AGN-cells are located inside forming clusters at a  high redshift, while at a  lower redshift only
one AGN-cell is usually found inside the virial volume. 
The shocks triggered by AGN feedback are efficient in removing the
cold gas phase at $z>1$, while producing
significant amounts of CRs in the forming cluster. Blast waves also cause the expulsion of a significant fraction of the hot gas from the cluster volume, in line with \citet{mcc2010}.

Figure \ref{fig:temp_cfr} shows a comparison of projected gas temperature across the entire AMR region of the same cluster, at three different cosmic epochs. 
While the large-scale morphology of the cluster is similar in all runs towards the end of the simulation, each feedback mode locally perturbs the ICM in a different way, driving powerful outbursts (or buoyant
bubbles) whenever strong cooling causes $n>n_{\rm min}$.

In the quasar-mode, the final large-scale morphology of the cluster is also affected by the overall action of powerful thermal bursts, and presents a large amount of ejected hot gas outside of the cluster volume.
In the jet-mode, prominent jets are launched along the horizontal axis, driving powerful shocks ($M \sim 5-10$) into the $\sim 10^6-10^7 \rm K$ cooling gas as well as gas motions
along their axis. The bubble-mode, on the other hand, has barely detectable effects on the large scale, and is globally similar
to a simple pure-cooling run (not-shown).

\begin{figure*}
\includegraphics[width=0.95\textwidth]{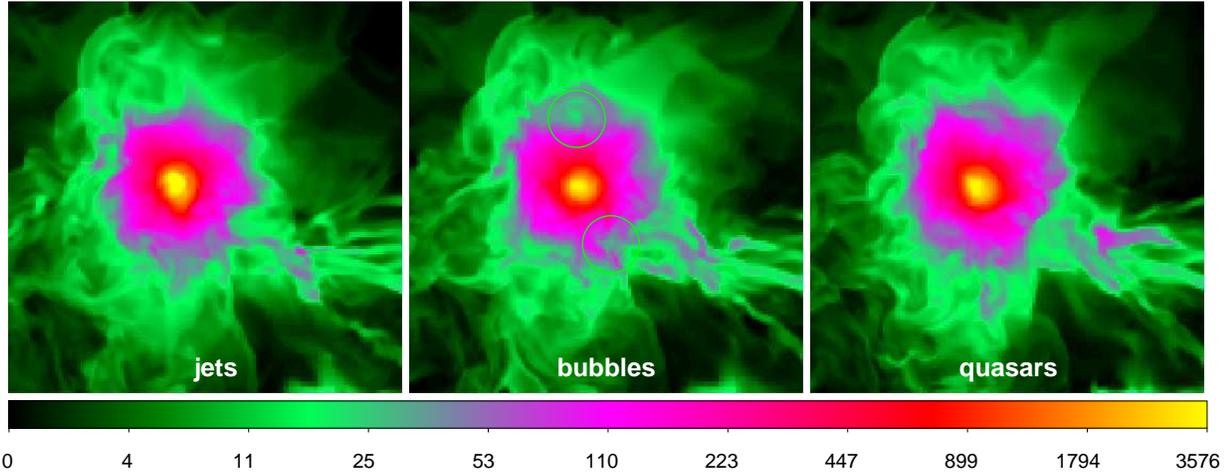}
\caption{Gas density (top panels) 
for a slice of $2 \times 2 ~\rm Mpc/h^2$ and width $25 ~\rm kpc/h$ across
re-simulations of cluster E5A at $z=0$. The color coding 
is [$\rho/\rho_{\rm cr,b}$] (where $\rho_{\rm cr,b}$ is the baryon critical density). 
The green circles show the projected location of two bubbles previously injected in the cluster centre.}
\label{fig:bubbles}
\end{figure*}

Fig. \ref{fig:bubbles} presents a zoomed image of the gas density maps for a slice of $2 \times 2 \rm Mpc^2$ and depth $25$ kpc$/h$ for the final configuration of cluster E5A in runs employing AGN feedback. Jets and quasars are not active any more at $z=0$ in this run, and no clear features related to their activity can be seen in the innermost cluster region at this redshift. A pair of bubbles, on the other hand, has recently been injected and can be seen as a couple of
under-dense blobs on opposite sides of the 
cluster centre (shown by green circles in the image).
Other images of active jets and bubbles can be seen in Fig. A1 of the Appendix.

\begin{figure*}
\includegraphics[width=0.99\textwidth]{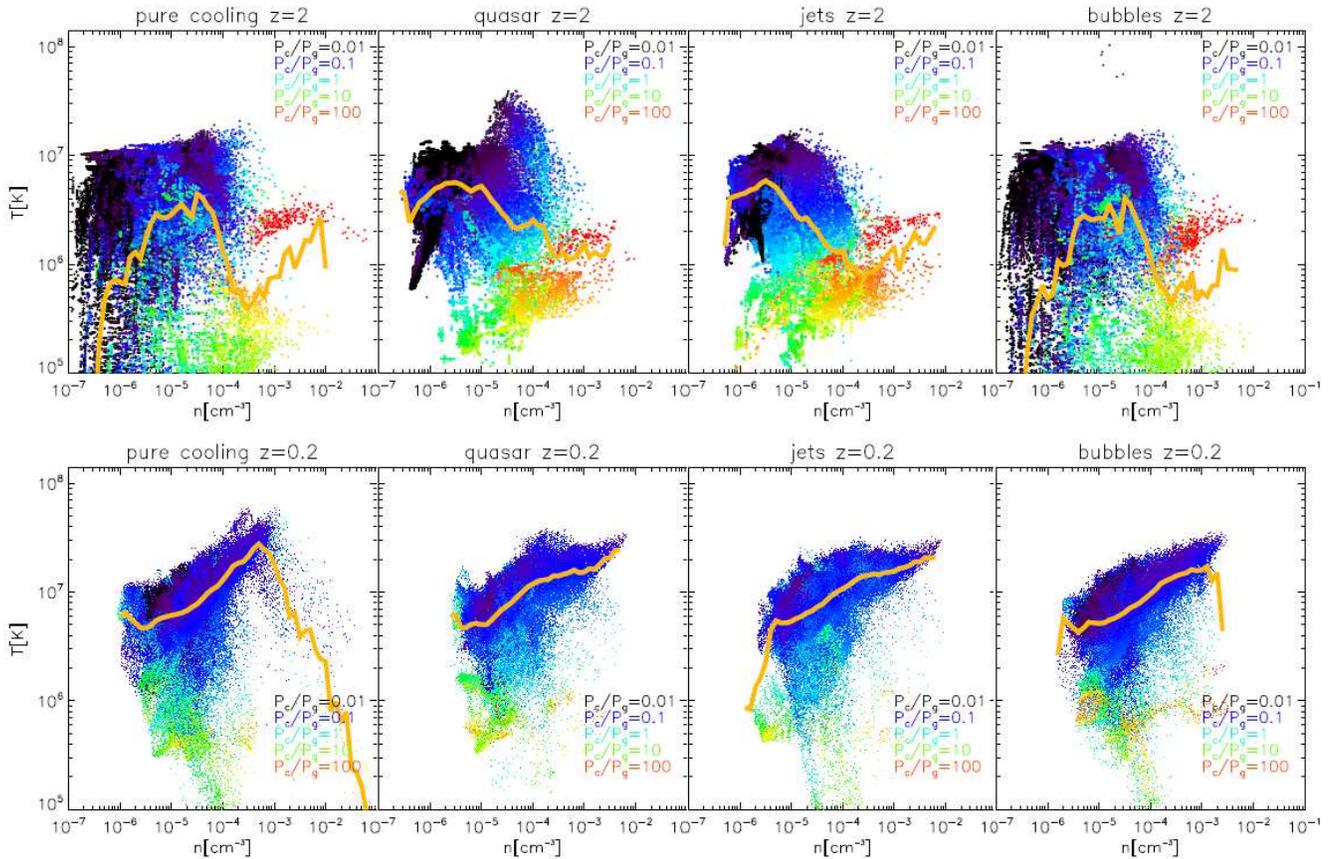}
\caption{Phase diagrams of cells within the virial volume of the re-simulations of cluster E1 at $z=2$ (top panels) and at $z=0.2$ (lower panels). The colour coding shows the cell-wise pressure ratio between CR and gas within cells, the thick line shows the average trend within each phase diagram.}
 \label{fig:phase}
\end{figure*}

We use phase diagrams for the cells inside the virial volume in order to characterize relevant differences. In Fig.\ref{fig:phase} we show the example of phase diagrams for an early cosmic epoch before the cluster formation, $z=2$ (top panels), and after the formation of the cluster, $z=0.2$ (lower panels) for the four different feedback modes.
The most evident difference between the runs is the absence of the "cooling flow" phase on the lower right part of the diagram, in all runs employing AGN feedback. In the bubble-mode, the feedback from buoyant bubbles only affects the high-density regions around the cluster core. At high redshift, the phase diagram is very similar to the pure cooling run, with the exception of the selective removal of gas with $n_{\rm min}>10^{-2} \rm cm^{-3}$ cells, which are heated to high temperatures ($>5 \cdot 10^{7} \rm K$) by the injection of bubbles.
 On the other hand, in the quasar and jet mode the action of feedback drastically alters the phase diagrams with respect to the
pure cooling run. This is particularly evident in the quasar mode at $z=2$, where a "cloud" of high temperature cells ($T>10^7 \rm K$ and $n \sim 10^{-4} \rm cm^{-3}$) is left after the passage of an AGN-driven blast wave.
At $z=0.2$, the "cooling" phase has almost disappeared in jet and quasar feedback, due to the efficient and volume-filling nature of the
heating events driven by such powerful mechanisms. In the bubble run  the phase diagram is similar to the pure-cooling run, with the exception of the $n \sim 10^{-2} \rm cm^{-3}$
regime, where bubbles are injected almost continuously and over-cooling
is balanced by mixing. This leaves a significant portion of 
"cold" ($<10^{6}$ K) and dense cells in the simulated volume, characterized by large values of $P_{\rm cr}/P_{\rm g}$.

\subsection{X-ray properties}
\label{subsec:xray}


We compare our simulations with 
the collection of {\small CHANDRA} cluster observations of \citet{cav09},  publicly available via the ACCEPT catalogue (http://www.pa.msu.edu/astro/MC2/accept).
The catalogue consists of 241 clusters  
mostly located in the redshift range $0.05 \geq z \geq 0.5$ and with average central temperatures
in the range $0.5-10 ~\rm keV$. For our comparison we selected a sub-sample of $\sim 60$ objects from the ACCEPT catalogue, with redshifts and average central temperatures compatible with our cluster dataset:
$z \leq 0.2 $ and $0.2 ~\rm keV \leq T \leq 5 ~\rm keV$ inside $0.2 R_{\rm 200}$.
For further details on the ACCEPT catalogue and set of observations, we refer the reader to \citet{cav09}.

\begin{figure*}
\includegraphics[width=0.9\textwidth]{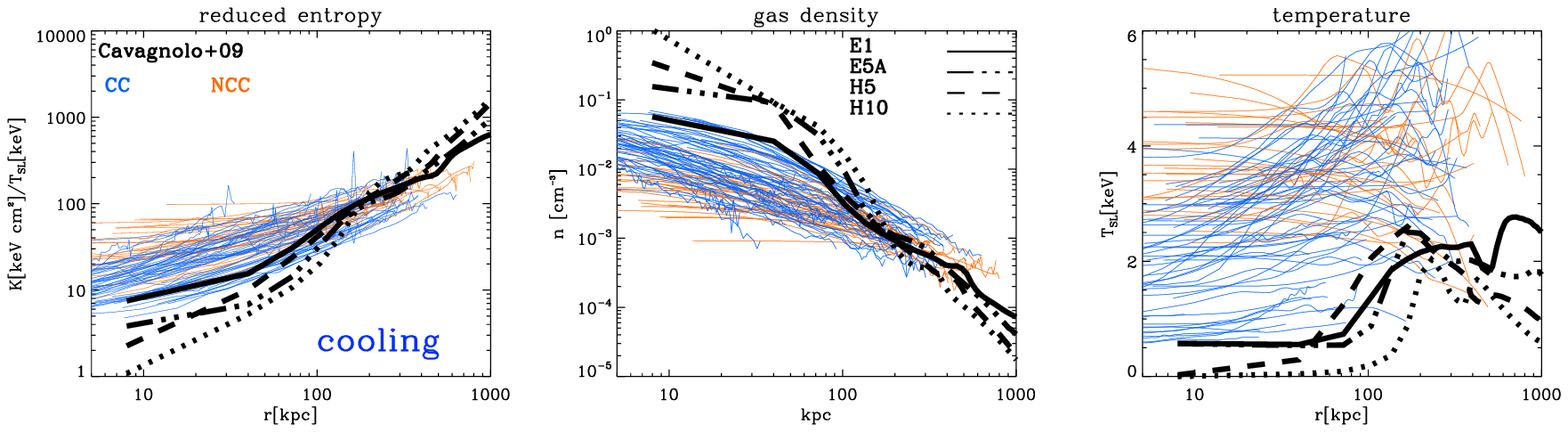}
\includegraphics[width=0.9\textwidth]{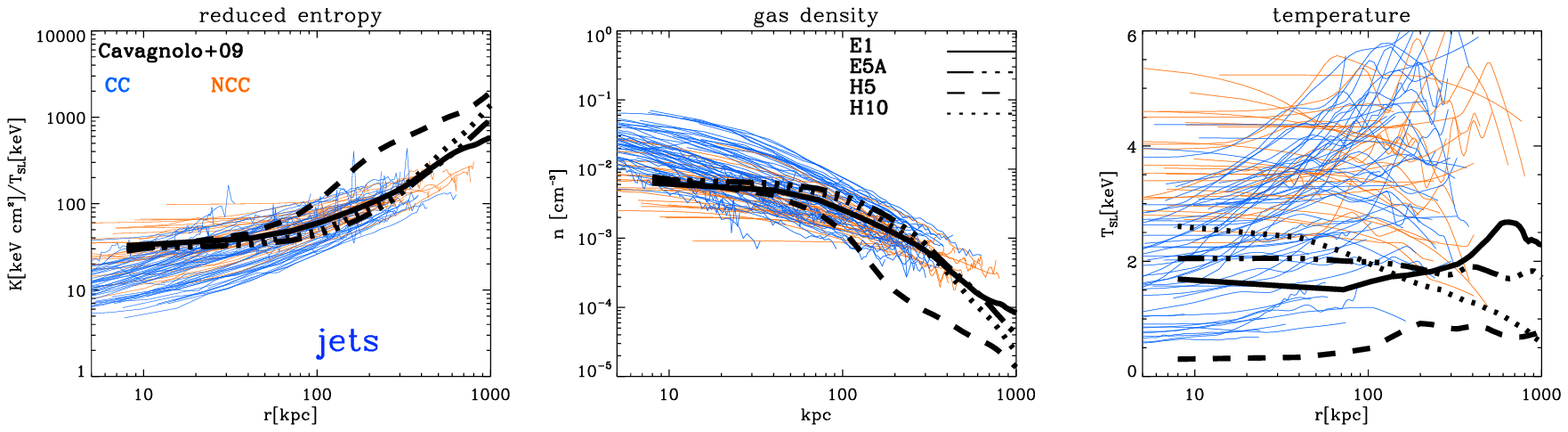}
\includegraphics[width=0.9\textwidth]{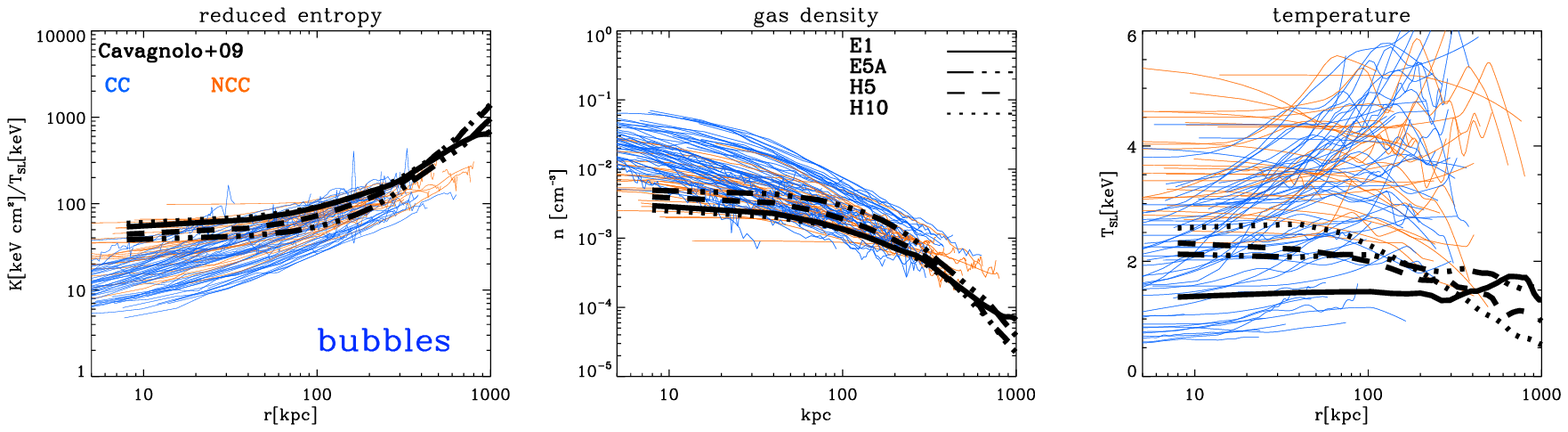}
\includegraphics[width=0.9\textwidth]{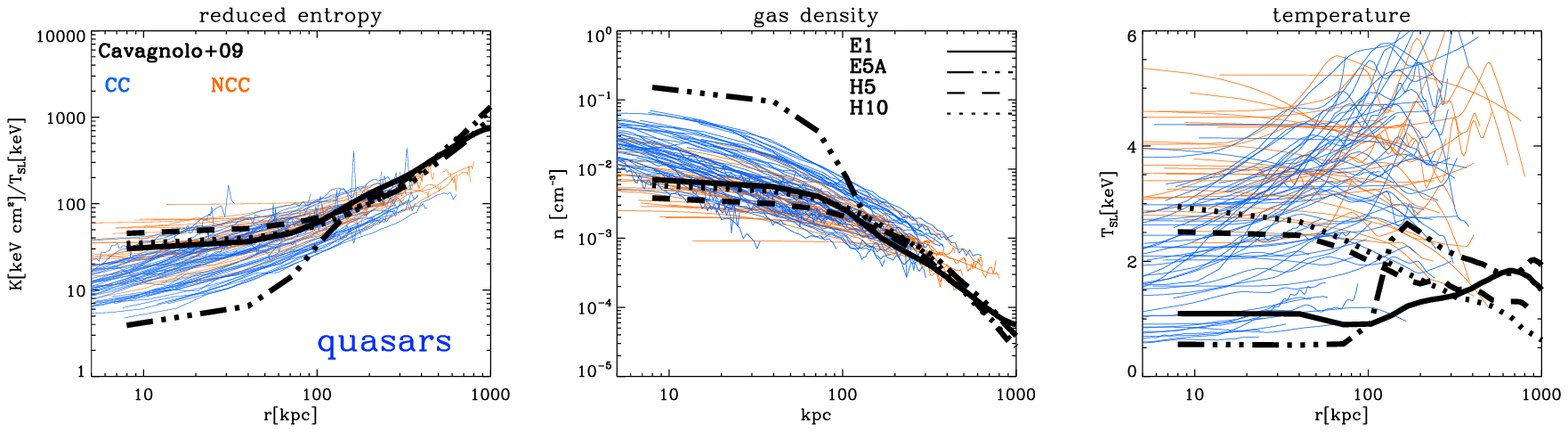}
\includegraphics[width=0.9\textwidth]{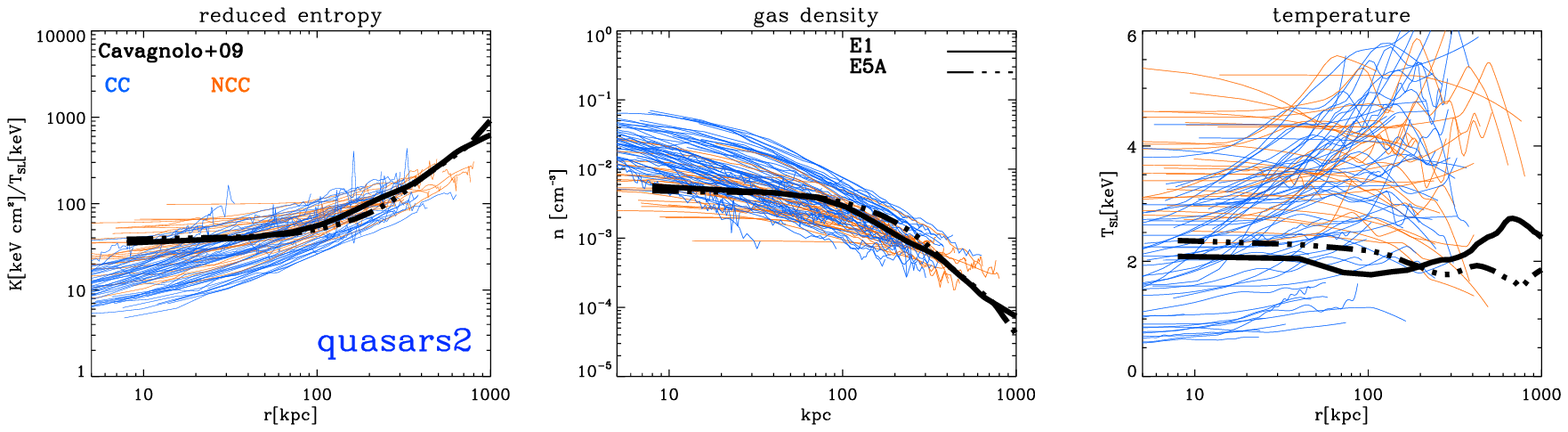}
\caption{Profiles of reduced gas entropy, gas number density and spectroscopic-like temperature for the simulated clusters (thick lines with different styles) and for observed clusters in the 
ACCEPT catalogue \citep[][the red lines are for non-cool-core clusters and the blue are for cool-core clusters]{cav09}. Each row shows the profiles of a different feedback
mode. The last row shows the profiles for the quasar mode, with $E_{\rm AGN}=10^{60} \rm erg$ per event (only the two most massive clusters are shown).}
\label{fig:profx}
\end{figure*}

In Fig.\ref{fig:profx}, we show the radial profile of reduced gas entropy {\footnote{We use the "reduced" entropy, rather than the usual entropy, $K=T/n^{2/3}$, since this removes the dependence on the host cluster mass, \citep{borgani08}.}}($K/T_{\rm SL}$, where $K=T_{\rm SL}/n^{2/3}$ and $T_{\rm SL}$ is the-spectroscopic like temperature, \citealt{2005ApJ...618L...1R}), gas number density, $n$, and spectroscopic-like temperature
for the different feedback modes. 
The simulated profiles are compared to the profiles of observed clusters, divided by the colour-coding into "cool-core" systems if their central entropy is $<30 ~\rm keV ~cm^2$ (in blue), or "non-cool-core" systems otherwise (red).

As expected, none of the pure cooling runs can reproduce the observed profiles. In these runs, steep gas density profiles with central values of 
$\sim 0.1-1 ~\rm cm^{-3}$ and reduced central entropy below $\sim 10 ~\rm cm^{2}$ are found, at odds with observations and in agreement with standard radiative runs \citep[e.g.][for a recent review]{2012ApJ...747...26L}.
Only cluster E1 is marginally within the range allowed by observations for density and entropy. However,
considering that this system has just gone through a strong major merger at ($z \sim 0.25$), it cannot be classified
as a classical cool-core cluster, as suggested by its gas 
density and temperature profile.
On the other hand, 
all runs employing AGN feedback yield a much better comparison with observations.
Jet and bubble feedback modes produce roughly similar profiles for all clusters: a rather flat
entropy profile within cluster cores (compatible with the high entropy floor of non-cool-core clusters
in \citealt{cav09}), a shallow density profile and an almost isothermal temperature profile inside the central $\sim ~\rm Mpc$ from the centre. 
Less clear results are found in the case of quasar-feedback. While in clusters E1, H5 and H10 a reasonable match with observations is found with the "fiducial" energy budget of $E_{\rm AGN}=10^{59} \rm erg$ per
event, this is insufficient to quench the cooling catastrophe in cluster E5A, where a rather
standard cooling flow takes place by the end of the simulation.
By increasing the available energy per event by one order of magnitude, $E_{\rm AGN}=10^{60} \rm erg$, 
the cooling flow is stopped also in cluster E5A (lower row of Fig.\ref{fig:profx}). Interestingly, the same is true for a re-simulation of cluster E1 using the same higher energy budget. 
However, in the case of the other smaller clusters the higher energy budget is too large and the thermal structure of both is destroyed, leading to a gas-poor cluster (see Appendix).

\begin{figure}
\includegraphics[width=0.45\textwidth,height=0.45\textwidth]{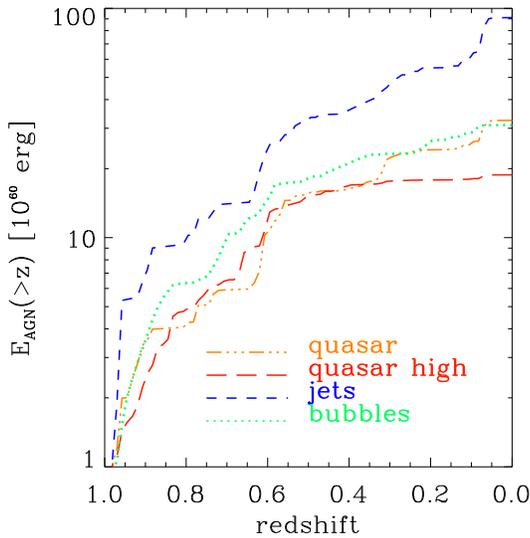}
\caption{Cumulative distribution of the AGN-energy released inside the comoving $(4 \rm Mpc/h)^3$ of cluster E1, from $z=2$ to $z=0$ for 4 investigated feedback models.}
\label{fig:tot_energy}
\end{figure}

This may simply suggest that different cluster masses must be characterized by different 
typical powers per event, mirroring the fact that a larger power per event is needed to balance the gravity and the pressure of the cooling gaseous atmosphere \citep[e.g.][]{2009MNRAS.395.2317P}.
However, it is interesting to notice that the self-regulating nature of AGN feedback (even with this simple implementation) yields very similar final profiles in the most massive object of our
sample, even if the power per event is 10 times larger.
In Fig.\ref{fig:tot_energy} we show the total energy released by AGN in the formation
region of cluster E1, in the redshift range $0 \leq z \leq 1$. By the end of the simulation, 
all feedback modes used a total amount of energy in the range $\sim 2-9 \cdot 10^{61} \rm erg$, corresponding to $\sim 0.01-0.09 $ of the total thermal energy of the cluster at $z=0$.

Interestingly, by the end of the simulation the higher power AGN mode ($E_{\rm AGN}=10^{60} \rm erg$) used about $\sim 30$ percent {\it less} energy than the lower
power run ($E_{\rm AGN}=10^{59} \rm erg$), which is  $\sim 1/3$ of the
total energy used in the kinetic feedback mode at a lower power. 
The total energy budget used in the bubble mode is similar to the quasar mode at the same power per event.
We note that in this system the use of AGN-feedback is significantly reduced
for $z>0.6$, due to the onset of large-scale mergers and powerful shock heating within the cluster volume. However, episodes of AGN-feedback
are present even at later epochs.  
Overall, a smaller amount of energy from AGN feedback is necessary to balance the catastrophic cooling with purely thermal feedback, or bubbles, compared to kinetic feedback. 
This is consistent with analytical results of \citet[][Fig.1]{2009MNRAS.395.2317P}, where it is shown that
a higher injection rate of kinetic energy, with respect to thermal energy, is necessary to balance the cooling flow for cluster masses $>10^{15} ~\rm M_{\odot}$. This is because in a massive cluster the critical momentum injection rate needed to overcome the pull of gravity and the surrounding gas pressure in the cluster core is more difficult to reach than the critical injection rate of thermal energy required to overcome the cooling flow via thermal feedback. In addition, the energy
release from jets close to the cluster core is very anisotropic, and it becomes
more isotropic (due to the driving of shocks) only at a distance of $\sim 100-200 ~\rm kpc$ from the core.

Overall, the X-ray properties of our clusters seem consistent with observed  low-temperature non-cool-core clusters since the typical internal drop in temperature of cool-core clusters is not observed. On the other hand, in the epoch in which the AGN feedback is triggered in our simulations, the thermal structure of the innermost cluster regions is not compatible with the appearance of typical cool-cores either, because the internal drop in temperature is typically much stronger than what is observed.  A lack of resolution compared to observations could be responsible for that.

However, we find no evidence of bimodality 
in the distribution of clusters entropy or gas density in our re-simulations, at odds
with observations \citep{cav09,2010A&A...511A..85P,eckert11}.
This might be connected to our coarse sampling of the parameters space of AGN feedback,
or may instead call for the implementation of more complex physics in our simulations.
However, the investigated sample is too small to address this important issue in detail, and we 
will leave this for future work.

\begin{figure}
\includegraphics[width=0.45\textwidth,height=0.4\textwidth]{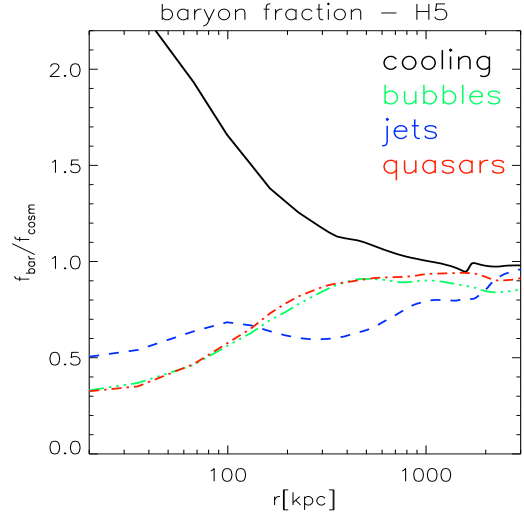}
\caption{Profiles of enclosed baryon fraction (normalized to the assumed cosmic value, $f_{\rm cosm}$), for the re-simulations of cluster H5 at z=0.}
\label{fig:prof_bar}
\end{figure}

We illustrate the effects of AGN feedback on the radial distribution of
baryon fraction by referring to the relevant example of cluster H5 at $z=0$, shown in Fig.\ref{fig:prof_bar}. As expected, the onset of the cooling flow causes too strong a concentration
of baryons inside the cooling radius,  clearly at odds with observation \citep[e.g.][and references therein]{2009A&A...501...61E,2012NJPh...14d5004S}.
On the other hand, all feedback runs produce a final distribution of gas mass which is more in agreement with observations, 
showing  increasing profiles towards
 the cluster outer regions  \citep[e.g.][]{2009A&A...501...61E,2012NJPh...14d5004S}.
Approaching $R_{\rm 200}$ the enclosed gas mass fraction is $90-100$ percent of the cosmological baryon fraction. However, the different implementations of feedback produce significantly different and time-dependent features in the profile. 
In particular, the mechanical action of jets changes the
shape of the enclosed baryon fraction in this low-mass system, due to a more recent feedback episode. Similar trends are found in
the other systems, provided that the time-dependent action of jets of this power has a decreasing impact on the shape of the baryon
fraction profile moving to larger masses.
 In the case of bubble-feedback, the artificial removal of gas when bubbles are created causes a small deficit of baryons inside the radius. Since the enclosed baryon fraction is very similar (within a few percent at all radii) to the one obtained with thermal feedback, where the gas mass is conserved instead, we conclude that the loss of baryon mass due to our procedure of generating bubbles is not the leading mechanism of the observed trend.

\subsection{Shocks and cosmic rays}
\label{subsec:shocks}
Blast waves triggered by explosive AGN feedback (in the quasar or jet mode) inject CRs at
$M>3$ shock waves in the ICM. This can happen before and after the formation of the cluster, as already shown
in Fig.\ref{fig:temp_evol_agn}. These shocks add to the underlying pattern
of cosmological shock waves driven by matter accretion, and to the
budget of shock-accelerated CRs that are continuously injected
into the ICM \citep[][]{ry03,pf06,sk08,scienzo}.

\begin{figure}
\includegraphics[width=0.495\textwidth]{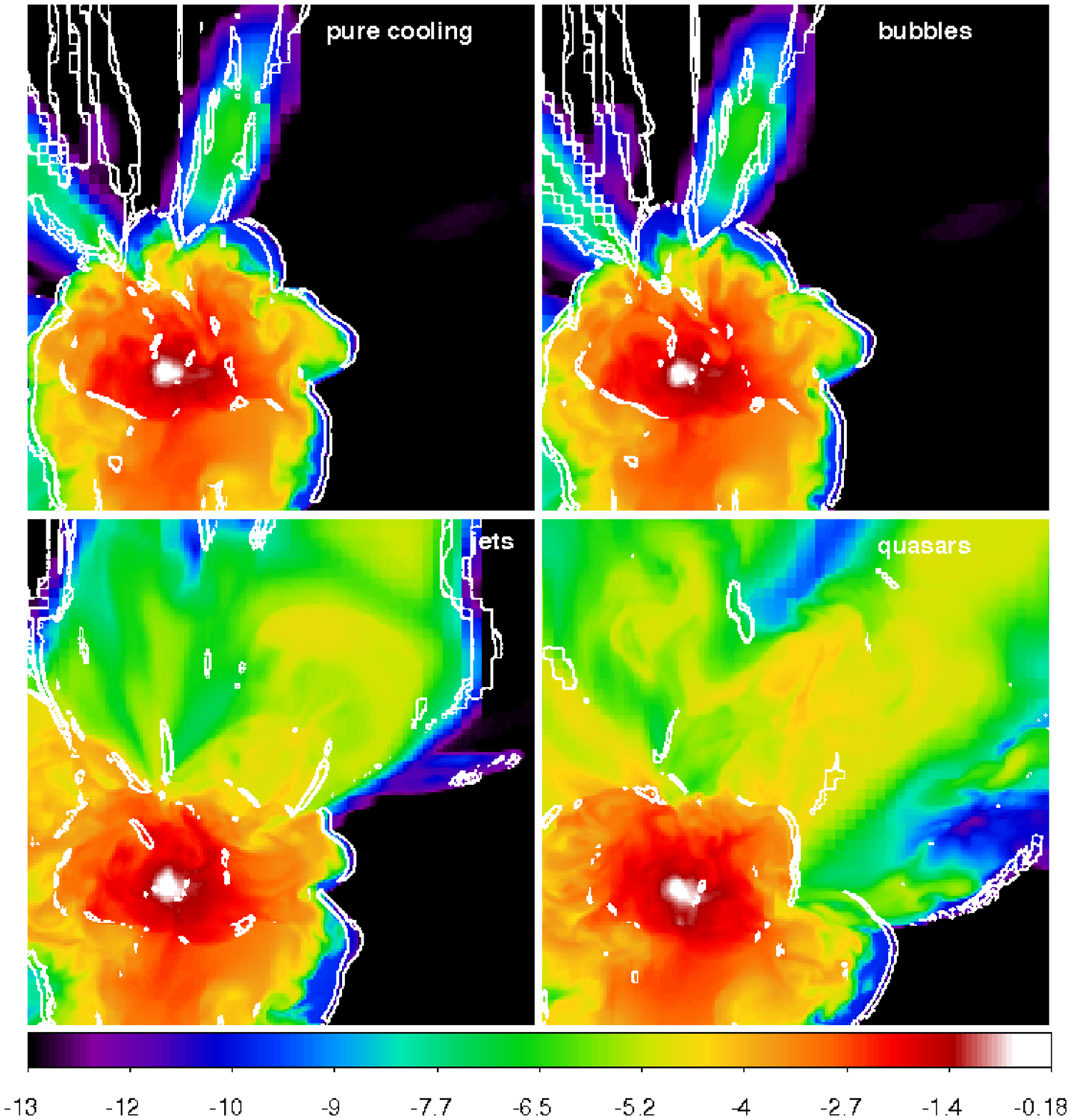}
\caption{Maps of average CR-energy for a slice of depth $100 ~\rm kpc/h$ in the four re-simulations of cluster H5 at $z=1$. Each panel has a side $5~\rm Mpc/h$ comoving. The colour coding is $\rm log_{\rm 10}(E_{\rm CR})$ (in arbitrary units), the overlaid colours show the location of $M>2$ shocks for the same region. }
\label{fig:ecr_mach}
\end{figure}

Figure \ref{fig:ecr_mach} gives an example of the different patterns
of shock waves and CR-energy in the re-simulations of one simulated cluster (we chose here the epoch of $z=1$, when AGN feedback is still very active). 
The pure cooling run and the run employing bubble feedback show a pattern of shocks
similar to non-radiative runs \citep[][]{scienzo}, with strong accretion shocks enveloping the cluster and
a few weaker substructure shocks inside the 
cluster. The CR-energy in these runs is concentrated within the cluster
and the filaments of matter being accreted onto it.
The re-simulations with jets and quasars present a much wider distribution of CR-energy, as a result
of previous episodes of gas and CR expulsion from the proto-cluster.
In this case, shocks launched by mergers and accretions add to those previously driven by AGN bursts. In general, along the directions of powerful outflows of gas and CRs the location of accretion shock from smooth material is shifted up to several $\sim \rm Mpc$ from the cluster center. This happens because the release of
non-gravitational energy heats the ICM along the 
outflows, and increases the sound speed there. 
Very similar trends are found also in the other clusters. 
This implies that, in general, the impact of feedback on the 
distribution and amount of CR energy can be significant not only 
inside clusters, but also in large-scale filaments associated with 
the cluster. 

\begin{figure}
\includegraphics[width=0.495\textwidth]{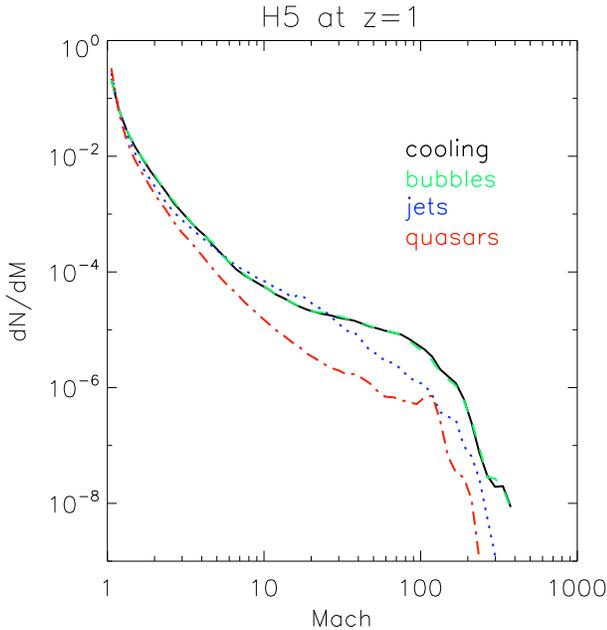}
\caption{Differential distribution of shock Mach numbers as a function of the AGN feedback mode, for the same simulated volume as in Fig.\ref{fig:ecr_mach}.}
\label{fig:dN_mach}
\end{figure}

The differential volume distribution of shock Mach number (measured as explained in Sec.\ref{subsec:cr}) for the same volume of $(5 ~\rm Mpc)^3$ of Fig.\ref{fig:ecr_mach} is shown
in Fig.\ref{fig:dN_mach}, for all feedback modes. 
At the epoch of this output ($z=1$), we find a systematic 
deficit of shocks in runs with quasar feedback, simply because
shocks have travelled out of the reference volume, and the ICM is overall much hotter with respect to
 pure-cooling or bubbles feedback. The run with jets displays
a distribution of Mach numbers closer to the pure-cooling case, but it shows a similar deficit of strong shocks inside the reference
volume. 

\begin{figure*}
\includegraphics[width=0.45\textwidth,height=0.4\textwidth]{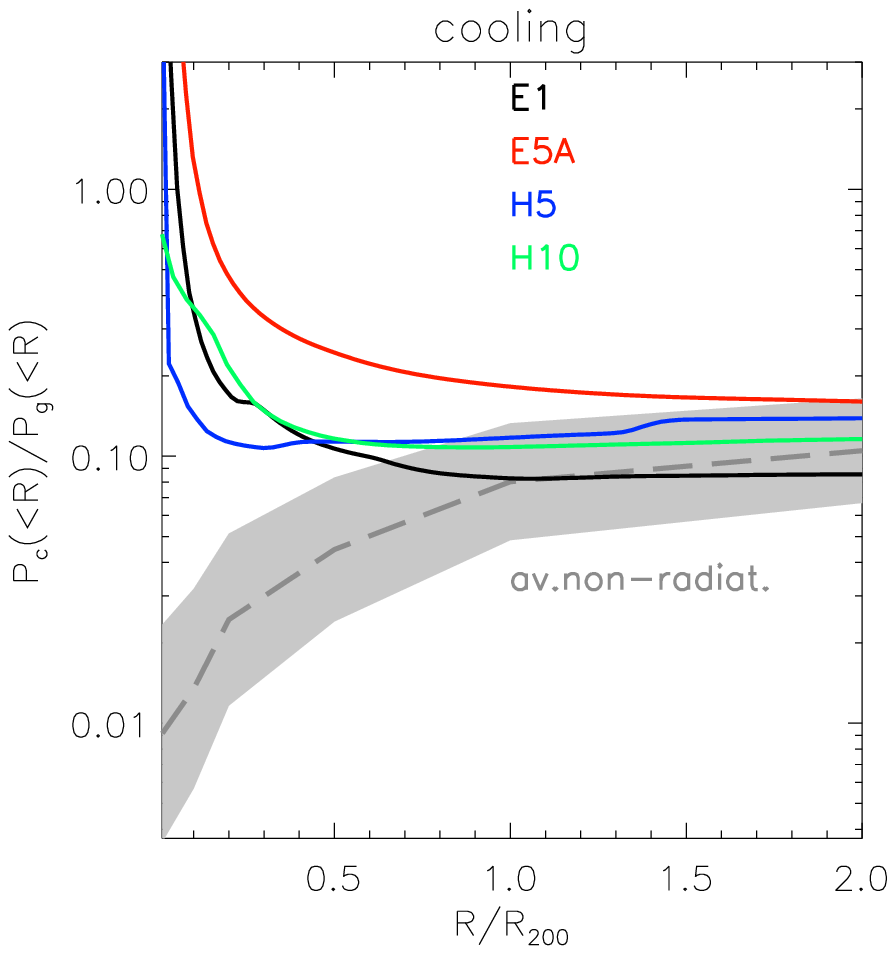}
\includegraphics[width=0.45\textwidth,height=0.4\textwidth]{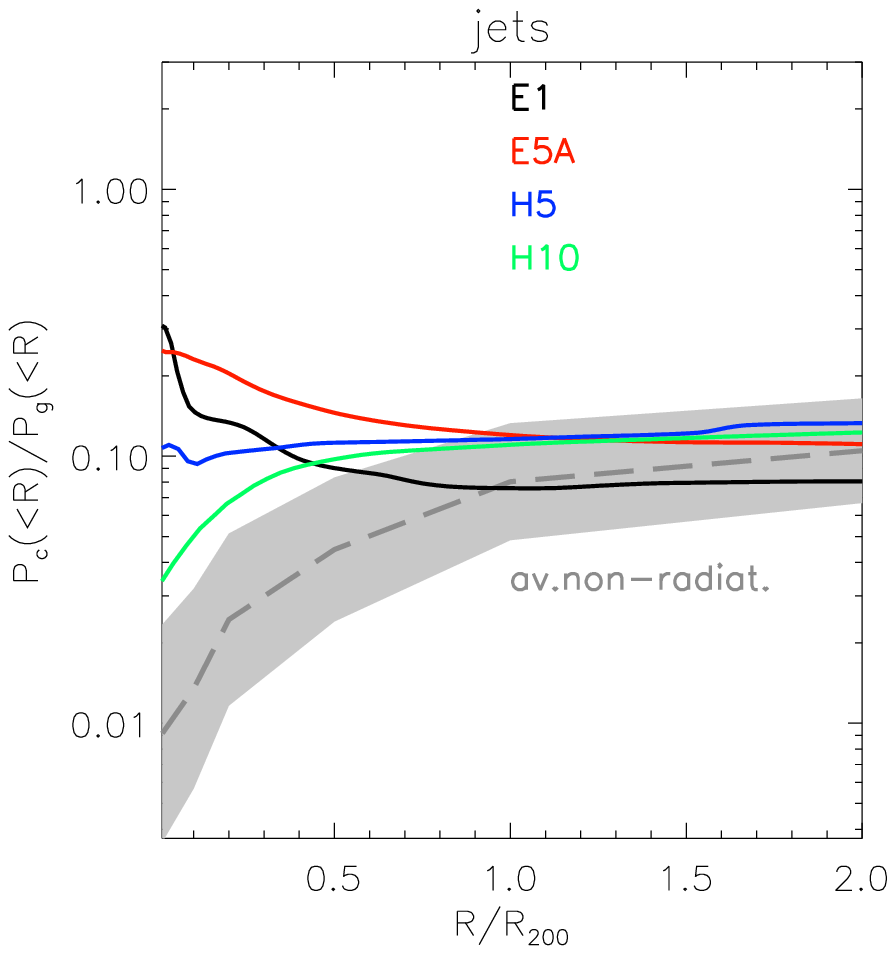}
\includegraphics[width=0.45\textwidth,height=0.4\textwidth]{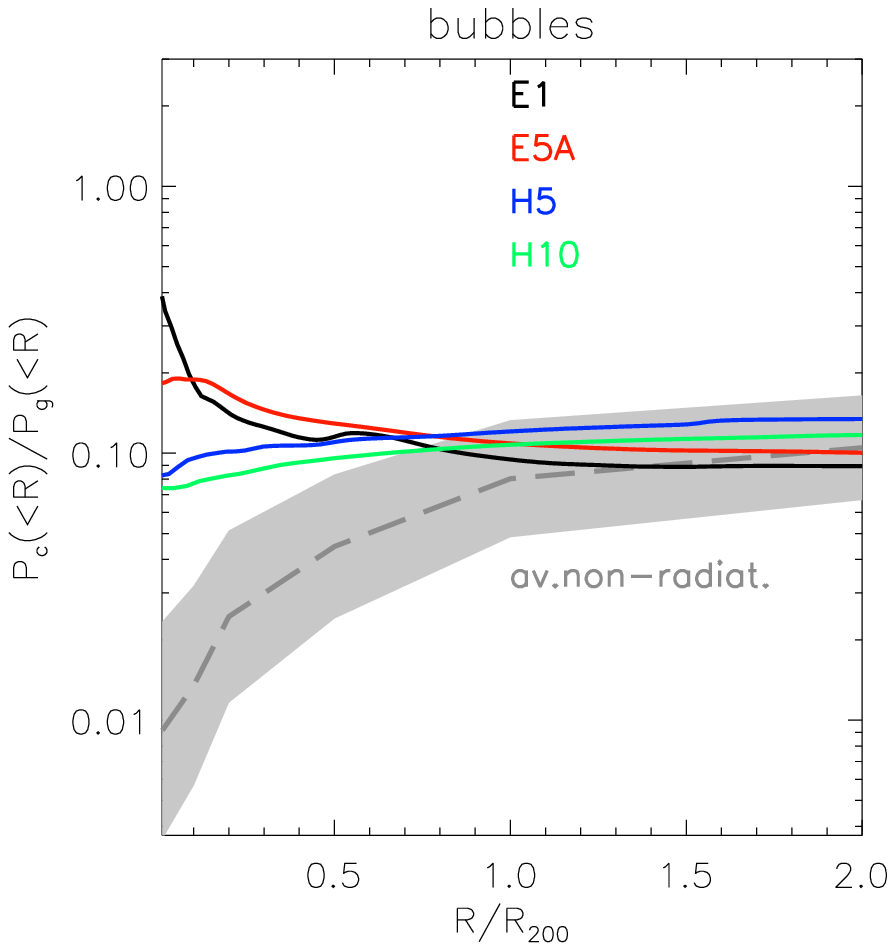}
\includegraphics[width=0.45\textwidth,height=0.4\textwidth]{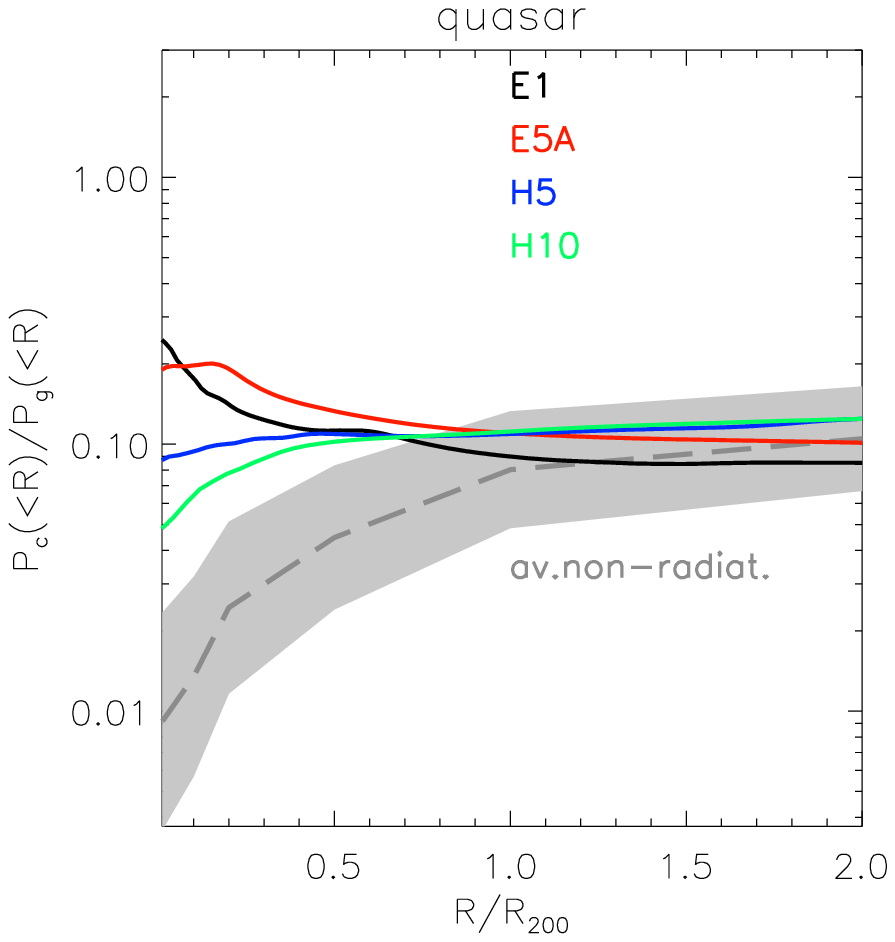}
\caption{Radial profiles of pressure ratio between CR and gas for our cluster
sample, in the case of pure-cooling (top left panel) or of different AGN-feedback
models. The shaded grey area gives the result (average and $\pm 1 \sigma$-scatter) of the non-radiative cluster sample of \citet{scienzo}.}
\label{fig:prof_pratio}
\end{figure*}

In Figure \ref{fig:prof_pratio} we show the profiles of the pressure ratio $P_{\rm cr}/P_{\rm g}$ for all clusters of the sample. 
As a reference, we also plot
the average pressure ratio obtained with the sample of 7 non-radiative cluster
runs presented in \citep[][]{scienzo}.

In all feedback modes we measure a ratio $P_{\rm cr}/P_{\rm g} \sim 0.1$
inside $R_{\rm v}$, within the range of what we already reported for
non-radiative runs employing only CR injection at cosmological shocks \citep[][]{scienzo}.
However, large variations are found by comparing pure-cooling or feedback runs.

Cooling acts to increase this ratio by removing gas from its hottest phase at all radii. In addition, pure-cooling is also found to dramatically increase the loss-rate of secondary particles in the centre, up to $\sim 10^{42} \rm erg/s$ (as shown in Fig. A2 of the Appendix). The heating term from CRs is, however, not sufficient to balance the radiative losses in the innermost ICM, which are usually $\sim 10-10^2$ times larger. Approaching the cluster core we measure $P_{\rm cr}/P_{\rm g} \sim 0.5-2$ in the smallest objects, and $\sim 1-10$ in the most massive ones. Unfortunately, our sample of clusters
is too small to allow us to infer a trend with the cluster mass. One possible reason for this trend can be, however, numerical: if the maximum spatial resolution is fixed (as in this case) the cooling region of largest
structures is better resolved, usually producing a slightly larger gas density peak and a significantly higher cooling rate. This trend in the
cooling rate has been reported in a number of works investigating SPH \citep[e.g.][]{2002ApJ...567..741V,2003MNRAS.342.1025T} or Eulerian simulations \citep[e.g.][]{1998ApJ...495...80B,2012ApJ...747...26L}.
This trend with radius of the pressure ratio is also similar
to what is found also with SPH codes \citep{2010MNRAS.409..449P,2012A&A...541A..99A}.

 We find that feedback acts to reduce the
pressure ratio $P_{\rm cr}/P_{\rm g}$ at all radii, this effect being larger
inside $R_{\rm 200}$. This suggests that, although AGN feedback on average triggers the injection of additional CR-energy in the cluster core compared to
simple radiative runs, its effect is that of reducing the dynamical
role of CRs, because the same process also
increases the thermal gas pressure. Although AGN feedback has
the effect of changing the innermost shape of the profiles of pressure ratio in
clusters H5 and H10 (producing a profile more similar to the 
non-radiative case, with $P_{\rm cr}/P_{\rm g} <0.1$ inside $R_{\rm 200}$), 
the profiles of the pressure ratio in clusters E1 and E5A still
show a maximum within the cluster core, with $P_{\rm cr}/P_{\rm g} \sim 0.2-0.3$.

This internal shape is opposite to the case of the pure-radiative runs, and seems to
be the ``imprint'' of the early cooling stage of both systems. We verified
that indeed this profile is already in place at earliest epochs
of cluster formation ($z \sim 1-2$) for both systems.
It seems that once that the ``cooling'' profile had enough time to form, its
imprint can hardly be erased by AGN feedback, especially in high-mass
systems where, for a given AGN-power, it is more difficult to produce
an efficient outflow of the CRs previously accumulated. 
Similar profiles of the CRs to gas pressure ratio have been reported for
single-object simulations of AGN feedback in cooling flow clusters \citep[][]{guo08,fuj11,fuj12}, suggesting
that this may be a stable feature of such physical models.
In order to study in detail the mass-dependence of $P_{\rm cr}/P_{\rm g}$ in cosmology, a much larger cluster sample is required.

In general, it is not straightforward to immediately relate the power released by feedback (as in Fig.\ref{fig:tot_energy}) to the injected CR-energy within the cluster volume. 
Shocks which follow the most recent event of jet or quasar feedback are surely responsible for the injection of new CR-energy inside the cluster, as shown in Fig.\ref{fig:ecr_mach}. This can be seen also in the close correlation 
between the power per event of AGN feedback and the resulting $P_{\rm cr}/P_{\rm g}$ pressure ratio inside clusters in the tests shown in Appendix (Fig. A3-A4).
However, the long-term evolution of the simulated ICM and of its balance between CRs and gas energy depends also on other factors, such as the density-dependent
secondary losses (Sec. \ref{subsubsec:hadronic}) and the further evolution
of matter accretion from the outer cluster region. Indeed, cooling and feedback
do not only affect the cluster core, but also the density distribution in 
the accreted substructures, causing the release of different pattern
of shock waves \citep[e.g.][]{ka07,pf07} and turbulent motions \citep[e.g.][]{valda11}. For this reason, it seems not possible to derive a simple prescription
to relate the relase of feedback energy to the budget of CR-energy within
clusters at low redshift, once that that all competing mechanisms are 
fully taken into account.

We can compare these values to the available constraints from observations (see also next 
Sec.\ref{subsec:gamma} for a close comparison with $\gamma$-ray observations). 
The present upper limits for a large sample
of nearby galaxy clusters observed with {\small FERMI} imply $P_{\rm cr}/P_{\rm g}< 0.05-0.1$ \citep[e.g.][]{ack10,jp11}, with a poor
dependence on the assumed spectra of CR. Also the limits on the presence of diffuse Mpc-scale
radio emission in clusters can be used to
constrain secondary electrons and thus the
energy density of the primary CR protons
\citep{br07,brown11}. In this case, the limits 
depend also on the cluster magnetic
field strength and are complementary to those
obtained from $\gamma$-rays. In the relevant case of
an average magnetic field in clusters of a $\sim \mu G$ \citep[][]{bo10},  
radio observations of clusters with no Mpc-scale radio emission
suggest that $P_{\rm cr}/P_{\rm g} \leq 0.05$ ($0.15$) in the case of
a spectrum with an index $\alpha=2.5$ ($\alpha=3$), while
the limits are less stringent for smaller average magnetic fields \citep[][]{br07,jp11}.
These limits usually refer to innermost
$\sim$ Mpc regions of clusters, where both
the number density of thermal protons and the magnetic field are
larger. At present no tight constraints are available for the clusters outskirts where the CR contribution might be larger.

According to Fig. \ref{fig:prof_pratio}, it seems that all runs with un-balanced cooling 
produce a ratio $P_{\rm cr}/P_{\rm g}$ that is incompatible with radio/$\gamma$-ray observations. In runs with AGN feedback, we find a consistency with the
above upper-limits in the two smallest size clusters, and a significantly
larger pressure ratio in the two largest systems (factor $\sim 2-3$ in the
center). However, when the $\gamma$-emission of our clusters is directly
compared to the available data from {\small FERMI}, the results of
AGN feedback are below the upper limits. This is because the $\gamma$-emission
is more sensitive to the density ($\propto n^2$) than to the pressure
ratio between CRs and gas ($\propto P_{\rm cr}/P_{\rm g}$), and therefore the net effect of AGN in our simulations is that of producing a lower $\gamma$-flux
compared to upper limits, although producing a large $P_{\rm cr}/P_{\rm g}$ in
the center.  In the case of radio upper-limits, the values of $P_{\rm cr}/P_{\rm g}$ in our runs can be better reconciled with radio data if the average spectrum of CRs is $\alpha \sim 3$ (or steeper). As we will discuss in Sec.\ref{subsec:gamma}, this is actually the case of most of the
CR-energy found in our cluster runs.

\subsection{$\gamma$-ray flux}
\label{subsec:gamma}

CR-protons colliding with thermal protons of the ICM produce secondary $\gamma$-ray emission
\citep[][]{1980ApJ...239L..93D}.
Since once accelerated CR hadrons can  
accumulate in galaxy clusters \citep{bbp97} and produce a long-lived $\gamma$-flux signature \citep{pe04,2003MNRAS.342.1009M,2008MNRAS.385.2243A,donn10,pi10}, observations in this energy range
can provide an important test-bed for feedback models. Indeed, feedback episodes producing similar X-ray features at $z=0$ might still
contribute a different enrichment
of CRs across a whole cluster lifetime, and therefore a different
$\gamma$-flux from our simulated objects.
In this Section we investigate the level of secondary emission resulting from the acceleration of CRs in the different modes of AGN feedback. 

In the two-fluid model used here, we cannot follow the spectral evolution of the CR population. Hence in order to compute the $\gamma$-ray flux from hadronic collisions we must assume a spectrum for the CRs at the time of observation.
Shocks which developed in the already formed dense and hot ICM ($z< 1$), largely dominate the energy budget
of shock acceleration in clusters, compared to earlier times ($z>1$). 
It is therefore reasonable that the resulting population of CRs at later epochs has a spectral index
that is associated with these Mach numbers.
Using SPH simulations modelling the injection and energy evolution 
of CRs at cosmological shocks, \citet{pi10} recently derived an average $\alpha=2.3$ injection spectrum for their
sample of simulated clusters (for protons in the range $\sim 1-10 ~\rm GeV$).
In our case, based on previous work with {\small ENZO}, we know that
shocks that are mainly responsible for the injection of CRs for $z<1$ are in the range $2.5 \leq M \leq 3.5$ \citep[e.g.][and references therein]{va11comparison}, corresponding to a particle injection spectrum of $\alpha \sim 3.3$. This follows from the basic relation

\begin{equation}
\alpha=2 \cdot \frac{M^2+1}{M^2-1},
\end{equation}

\noindent that applies to the case of non-radiative shocks.

In the following we will base our computation of the $\gamma$-flux on the limiting cases of $\alpha=2.5$ ($M=3$) and $\alpha=3.3$ ($M=2$). 

Within the virial volume of clusters, the $\pi^{0}$ decay
is expected to dominate the cluster emission above $>0.1 ~\rm GeV$ with respect to other radiative mechanism (e.g. Inverse Compton of secondary electrons, non-thermal bremsstrahlung, \citealt[][]{2003MNRAS.342.1009M,2008MNRAS.385.2243A}). 
We also neglect turbulent re-acceleration of CRs \citep[e.g.][]{br07,bl11b}.
{\footnote{We notice that in principle also the signal from Dark Matter annihilation can contribute to the same energy range $[1-10] ~\rm GeV$ \citep[e.g.][]{2011PhRvD..84l3509P}. However, this has not been observed yet \citep[e.g.][]{2012ApJ...747..121A}, and the effect obviously strongly depends on the assumed cross-section of DM, a still open topic, and we do not model this contribution in any way. Indeed, our simulations can be helpful in assessing the statistical emission floor of $\gamma$-emission expected from structure formation processes and AGN feedback.}}
Following the formalism of  \citet{pe04} (see also \citealt{donn10}), we compute the  emission from an asymptotic
spectrum of particle, with a spectral index $\alpha_{\gamma} \approx \alpha$. 
The threshold proton energy is set to $E_{\rm min}=1 ~\rm GeV$.

\begin{figure}
\includegraphics[width=0.49\textwidth,height=0.22\textheight]{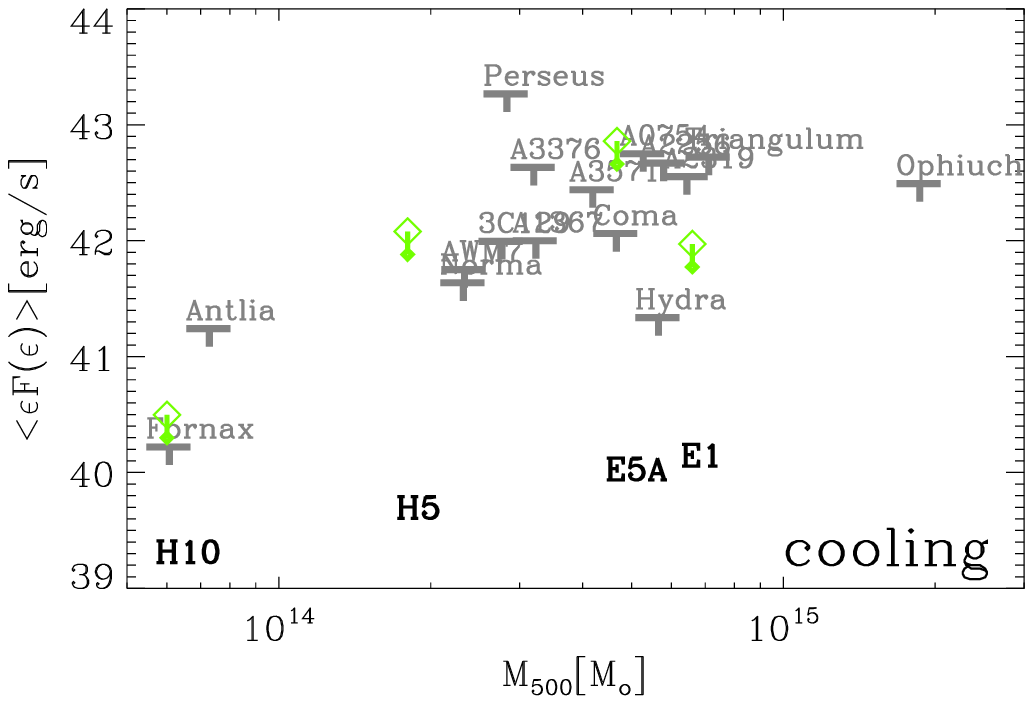}
\includegraphics[width=0.49\textwidth,height=0.22\textheight]{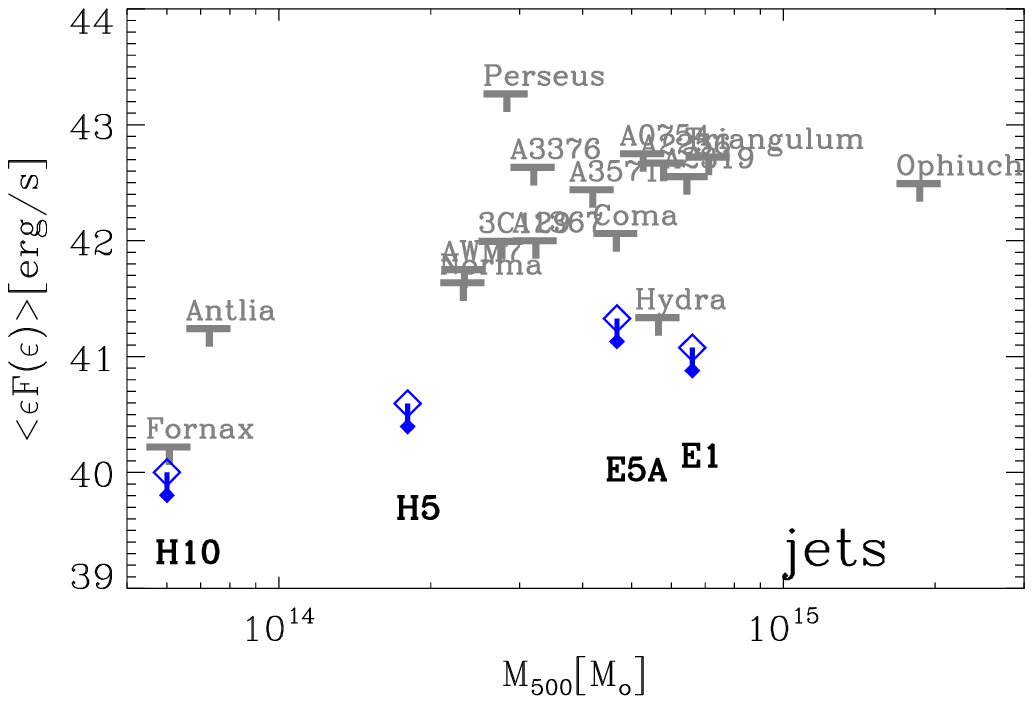}
\includegraphics[width=0.49\textwidth,height=0.22\textheight]{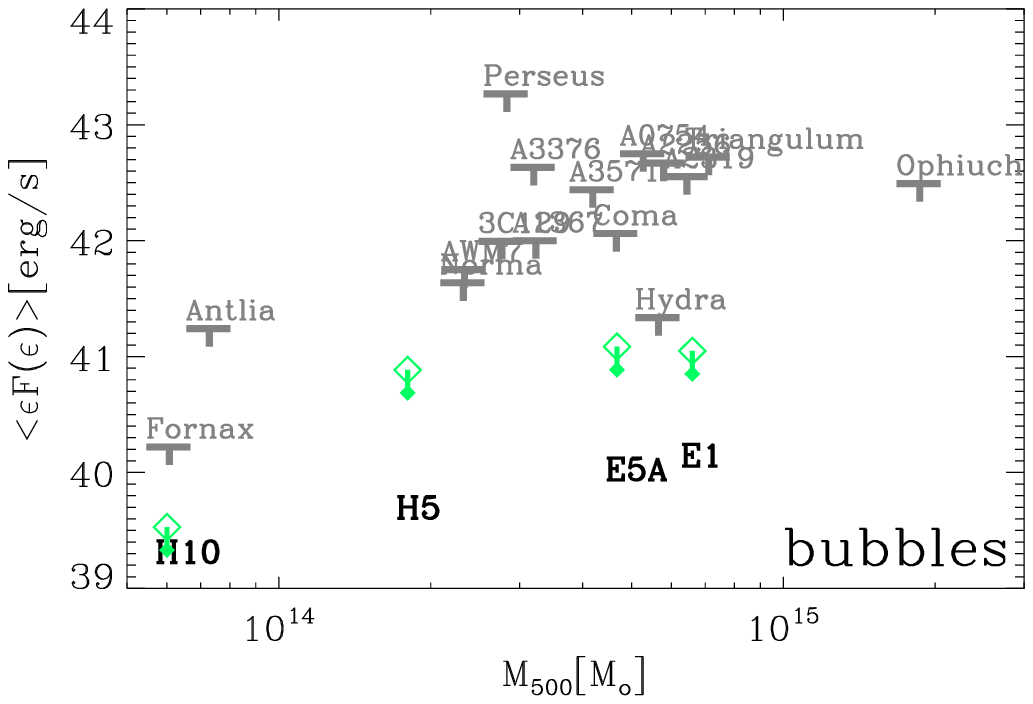}
\includegraphics[width=0.49\textwidth,height=0.22\textheight]{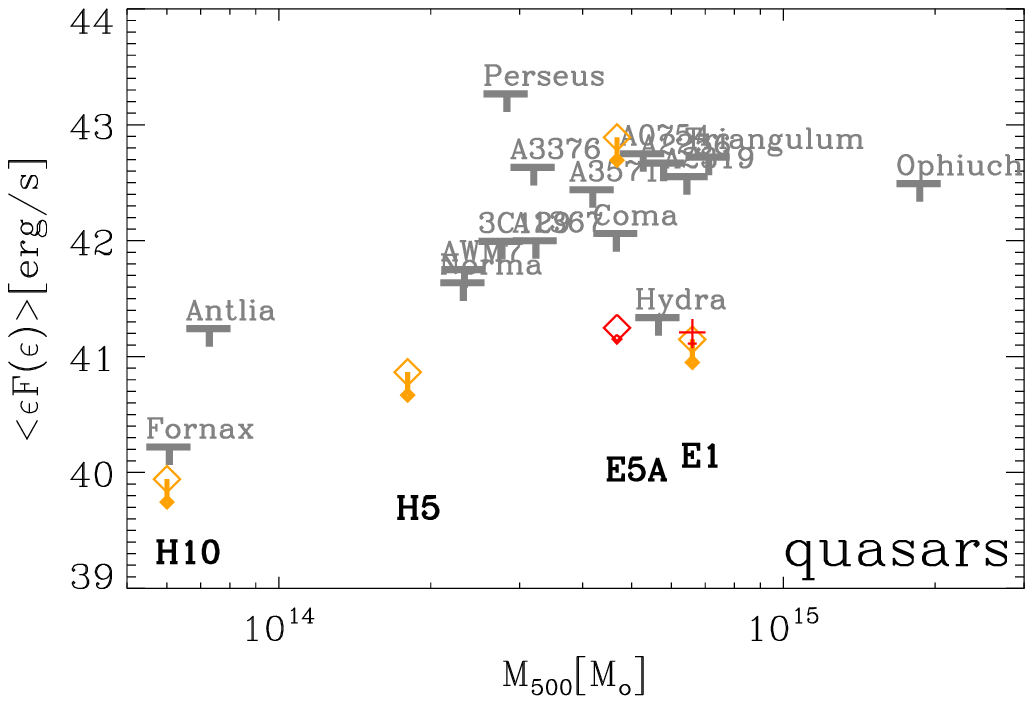}
\caption{Simulated $\gamma$-ray luminosity resulting from proton-proton for the $1-10$ GeV energy range, as a function of $M_{\rm 500}$. The large symbols are for $\alpha=2.5$ spectra and the small symbols are for $\alpha=3.3$ spectra. 
In the bottom panel (quasar mode), we additionally show with red symbols the result from the re-simulations of cluster
E1 and E5A employing $E_{\rm AGN}=10^{60} \rm erg$ per event. In each panel we compared the simulated data to the available
upper limits from {\small FERMI} \citep[from][]{ack10}. }
\label{fig:gamma}
\end{figure}

Our results for each implementation of AGN are shown in Fig.\ref{fig:gamma}, where we show the absolute luminosity of simulated clusters in the $[1-10] ~\rm GeV$ energy range,
(typically the most sensitive in terms
of average differential energy flux)
 and the upper limits
from {\small FERMI} observations of \citet{ack10}, as a function of the total cluster mass within $R_{\rm 500} \approx R_{\rm vir}/2$. 
The results in this
energy range do not depend much on the assumed spectrum of particles,
and the difference between $\alpha=2.5$ and $\alpha=3.3$
is only $\sim 30$ percent in $\gamma$-flux (shown by different symbols). 

All pure cooling runs produce an excess of $\gamma$-emission compared to
{\small FERMI} upper-limits, up to a factor $\sim 10$. This is mainly due
to the large density of thermal targets for the proton-proton collision
within the cooling radius. 

Runs with AGN feedback, on the other hand, are  below present-day {\small FERMI} upper limits (with the exception of the quasar run of E5A, with $E_{\rm AGN}=10^{59} \rm erg$ per event).
The bubble-feedback seems to provide the lowest amount of $\gamma$-emission across the sample, with $\gamma$-ray flux about $\sim 10$ times below the
{\small FERMI} upper limits within each mass bin. Jet and quasar feedback also produce $\gamma$-emission below the upper limits, but by a smaller factor.

Based on current upper-limits from {\small FERMI} \citep[][]{ack10} and the rather
small sample of simulated clusters, it is not possible to
reject any of the feedback models based on the
predicted $\gamma$-ray luminosity.
However, this approach can put robust constraints on feedback models and energetics, complementary to the analysis of X-ray profiles. Indeed, while the radial profiles of
gas entropy, temperature and density studied in Sec.\ref{subsec:xray} cannot alone lead to a clear assessment of which feedback mode is more suitable to explain observations, a $\sim 5-10$ times lower limit on $\gamma$-emission could already rule out the jet/quasar modes.

\citet{fuj12} recently simulated the effects of injection, streaming and heating of CRs from the
central AGN in the Perseus clusters, producing estimates of
non-thermal radiation from secondary particles at several
wavelengths. The $\gamma$-ray luminosities produced by
our AGN runs for cluster H5 (which has a mass close to Perseus)
are of the same order of magnitude as in the fiducial model in \citet{fuj12}, $\sim 10^{41} \rm erg/s$. This is $\sim 2$ orders of magnitude below the {\small FERMI}
upper limit at $[1-10] ~\rm GeV$.
Our AGN runs are similar 
to other cosmological simulations that estimate
the $\gamma$-ray power from secondary particles in clusters \citep[][]{donn10,2010MNRAS.409..449P}, our results.
Contrary to \citet{2010MNRAS.409..449P}, we do not excise dense overcooled clumps 
from the simulated volume and the contribution of shocks related to
AGN activity is added to the contribution of CRs from cosmological shocks. While a part of the gas mass locked into these dense clump would eventually lead to star formation and produce cluster galaxies, these structures are the same that are responsible for the production of the un-realistic X-ray properties (Sec.\ref{subsec:xray}) and therefore their excision would just ``mask'' a dramatic problem of the simulation. In addition, without the inclusion of star-formation processes in our simulation, it is difficult to establish which part of these
cold clumps would turn into galaxies, and therefore removing their contribution in $\gamma$-emission would be arbitrary.
The $\gamma$-ray power from the CRs accelerated at cosmological shocks
only is slightly smaller than
in \citet{2010MNRAS.409..449P}. The fact that the total $\gamma$-ray power from our clusters is lower than in SPH calculations stems from the typicall lower 
innermost gas density resulting in Eulerian simulations, and also in the 
the different way in which SPH and grid methods model the accretion
of gas matter within cluster cores, as already investigated
in \citet{va11comparison}. For the numerical reasons for that, we
refer the reader to \citet{ag07,wa08,mi09}.

\begin{figure*}
\includegraphics[width=0.33\textwidth,height=0.66\textwidth]{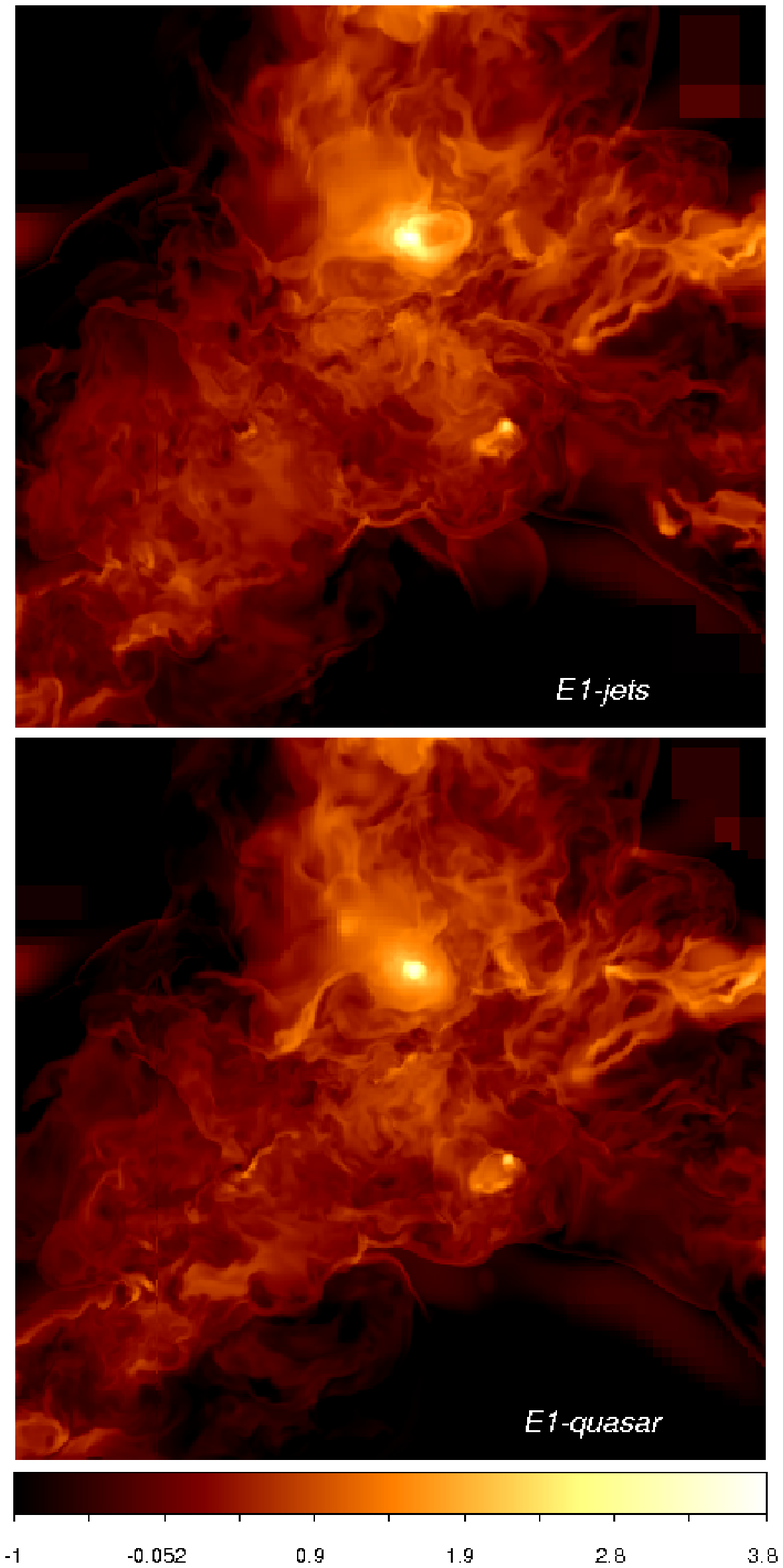}
\includegraphics[width=0.66\textwidth,height=0.66\textwidth]{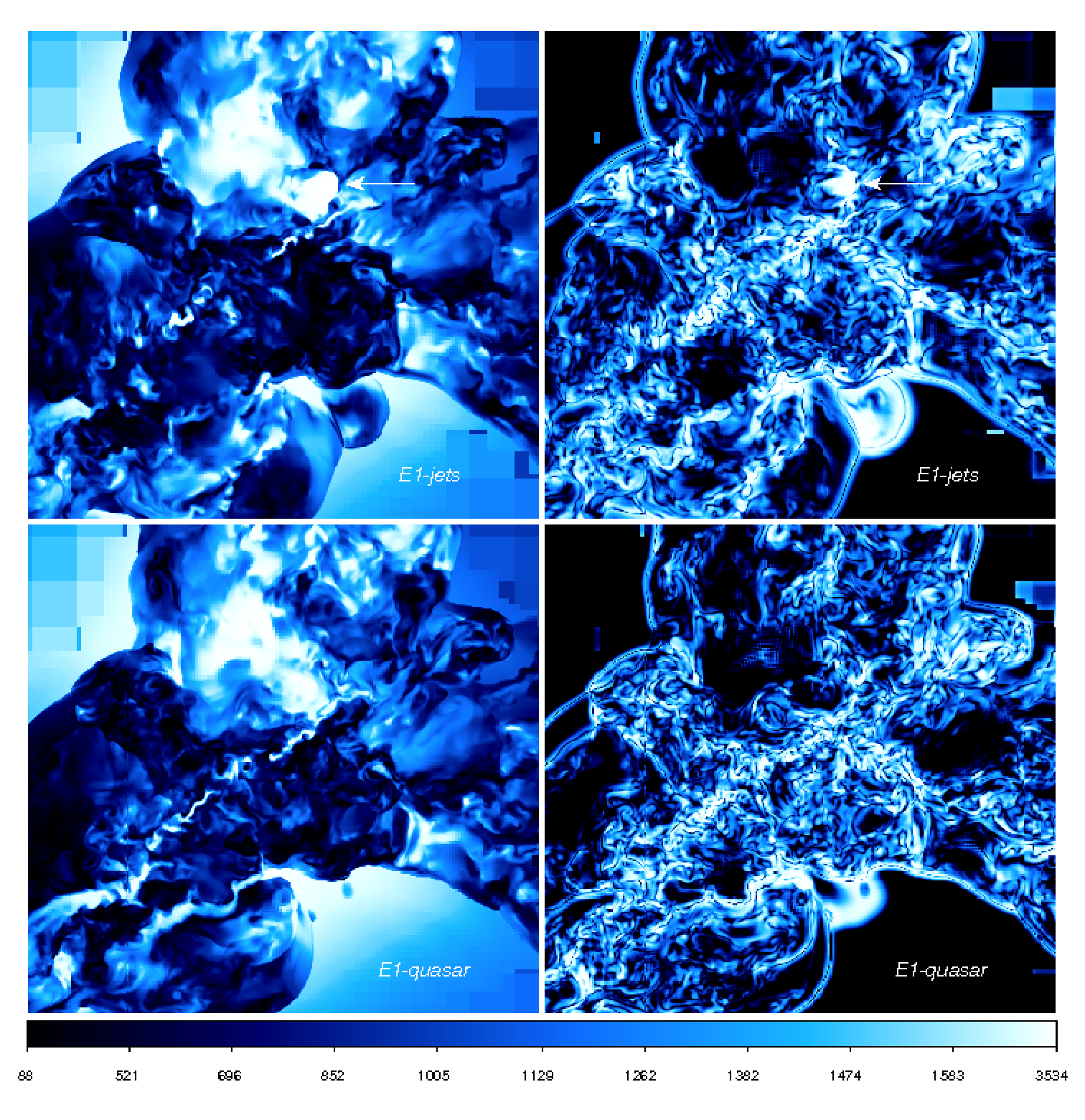}
\caption{Maps for a $4 \times 3 ~\rm Mpc/h^2$ slice of depth $25 ~\rm kpc/h$ for
cluster E1 at $z \sim 0.2$, simulated with jets (top panels) or quasar (bottom panels). The quantities shown are gas density (first column, $log_{\rm 10}$[$\rho/\rho_{\rm cr,b}$]), total gas velocity (central column, [km/s]) and the turbulent
gas velocity (last column, [km/s]. Note that the turbulent velocities have been multiplied by $5$ in order to use the same colour bar of the total velocity field). The horizontal arrows in the jet-run show the location of a recent
jet event.}
\label{fig:turbo_fig}
\end{figure*}

\subsection{Turbulent motions}
\label{subsec:turbo}

Feedback from AGN may add turbulent energy to the already turbulent 
state of the ICM. Unlike in our previous work \citep{va12filter}, with this set of simulations we
can study turbulent motions from, both, the accretion of matter and AGN feedback
(in its different modes).

In Figure \ref{fig:turbo_fig}, we show a map of turbulence reconstructed with the same multi-scale filtering technique of \citet{va12filter}, considering one run with jet-feedback (top panels) and in one with quasar-feedback (bottom panels). From left to right, the maps of gas density, total velocity and turbulent for a thin slice ($25 ~\rm kpc/h$)
through the cluster centre are shown. 
This system has an ongoing merger at this redshift, plus large-scale accretion of gas/DM matter along at least three filaments. 
These processes drive motions with bulk velocities of $\sim 1500-2000 ~\rm km/s$ on
$\sim ~\rm Mpc$ scales, and turbulent velocities of $\sim 600-800 ~\rm km/s$ on
scales of $\sim 300 ~\rm kpc$. In the run with jets an additional horizontal 
velocity pattern of $\sim 600-700 ~\rm km/s$ is present (indicated by an arrow within the panels). This is the outcome of a recent jet burst, launched $\sim 0.1 ~\rm Gyr$ ago. Even if the jet output dominates the velocity and the turbulent
velocity field in the surroundings of the cluster core ($<100-200 ~\rm kpc$), making
it anisotropic, the overall velocity and turbulence within the cluster volume
are dominated by the large merger, with a pattern similar
to the quasar-mode.

In general, we find that feedback affect the turbulent velocity field in our
simulations in two ways. First, by driving turbulence locally to the
AGN (e.g. along the jets or behind the wake of bubbles), in addition to the overall cluster velocity field. This is a general findings in simulations \citep[e.g.][]{2008ApJ...686..927S,heinz10,gaspari11b,dubois11}.

Second, by affecting the compactness of accreted substructures and cluster satellites, thus affecting their ram-pressure-stripping
and their efficiency in stirring the ICM. 
Overall, this makes the trend of turbulence in the different feedback modes non-trivial to identify, and close to the AGN highly time-dependent.

\begin{figure*}
\includegraphics[width=0.32\textwidth]{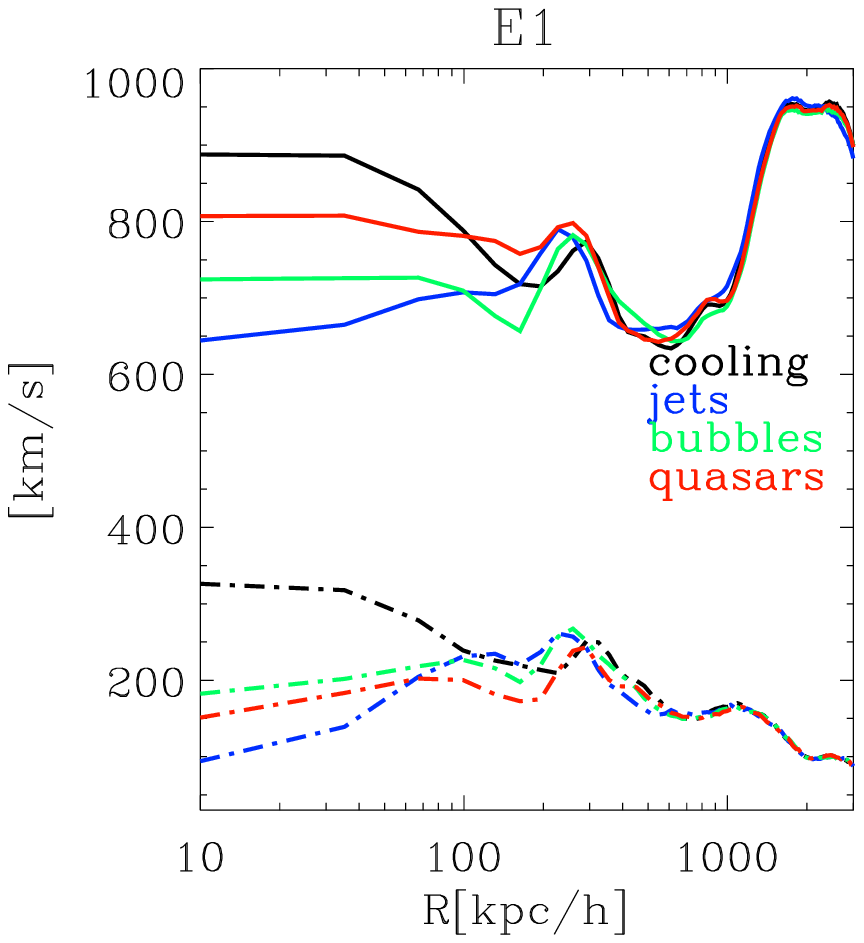}
\includegraphics[width=0.32\textwidth]{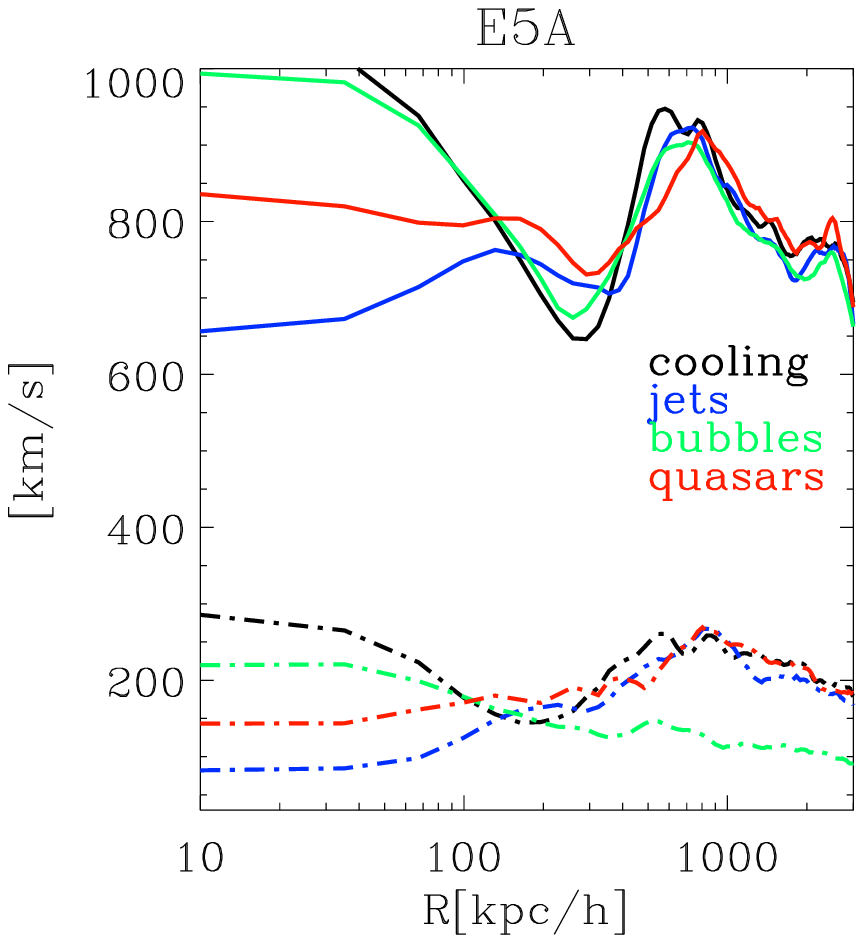}
\includegraphics[width=0.32\textwidth]{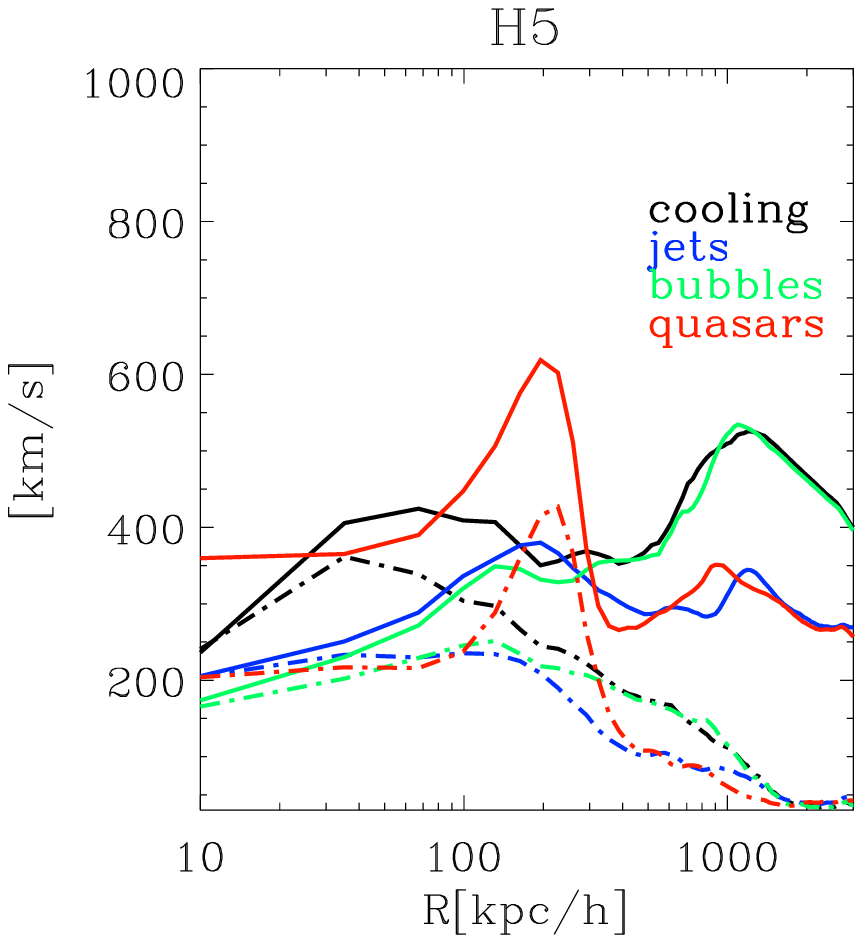}
\caption{Radial profiles of total velocity (upper lines) and turbulent
velocity (lower lines) for different re-simulations of E1, E5A and H5.}
\label{fig:turbo_prof}
\end{figure*}

This is confirmed also by the radial profiles of velocity and turbulence for the three most massive clusters (Fig.\ref{fig:turbo_prof}).
While the velocity profiles at $z=0$ are quite similar (except that for the cluster
core) in E1, increasingly larger variations between modes are found in E5A and in H5.
In all cases,  the pure cooling run produces the highest total
and turbulent velocity field. This is typical of
pure-cooling runs \citep[e.g.][]{dubois11,2012ApJ...747...26L}, 
and a result of the transonic
cooling flow. 
In addition, overcooling leads to more compact substructures, that are
more efficient drivers of turbulent motions during their orbits \citep[e.g.][]{valda11}.

\citet{sa11} recently computed upper limits from the maximally allowed amount of turbulence for a collection of 28 nearby (mostly cool-core) galaxy clusters, by fitting a thermal multi-temperature spectrum to observed {\small XMM-Newton} spectra. Following the same procedure as in \citet{va11turbo}, we computed the turbulent velocity field  on  $\sim 30$~kpc scales (which roughly corresponds to the projected volume available to the observations of \citealt{sa11}) for our cluster runs.
Since the turbulent
energy spectrum on the smallest spatial scales may be affected by numerical effects on the smallest scale
\citep[e.g.][]{pw94}, the energy of the turbulent motions on scales $ \leq 30$~kpc was calculated analytically from the measured total power spectrum on larger scales assuming Kolmogorov scaling {\footnote {In this case the Kolmogorov slope maximises the possible contribution of unresolved structures of the velocity field, since in general a slightly steeper power-spectrum is measured in these simulations at the smallest scale \citep[][]{va09turbo,va11turbo}. Numerical 
dissipation makes it difficult to constrain the slope at the smallest scale with
precision, and therefore our choice here already provides the most robust possible
test against observed upper-limits of turbulence.}}

\begin{figure}
\includegraphics[width=0.45\textwidth]{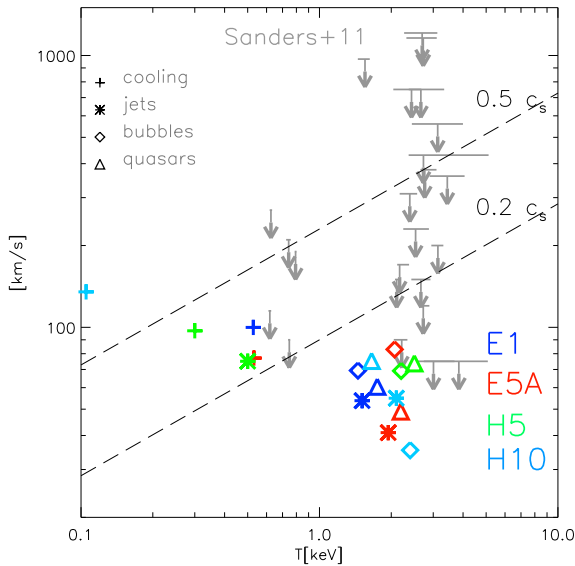}
\caption{Scaling between the average temperature and the mean velocity dispersion (extrapolated for $<30 ~\rm kpc$) for all cluster runs, and for the observational sample by \citet{sa11}. The additional dotted lines show the dependence of the ICM sound speed with the temperature. 
}
\label{fig:sanders_limits}
\end{figure}

In Fig.\ref{fig:sanders_limits} we show the comparison between the simulated
points at $z=0$ and the {\small XMM-Newton} limits for the galaxy clusters of the sample. 
As in \citet{va11turbo}, the values for the cores of simulated clusters are generally below the observed upper limits. However, the pure-cooling
runs show a velocity slightly in excess of some observed objects. AGN feedback generally reduces turbulent velocities within the
core (as shown above) and pushes the observed temperature from
the core to higher values. Basically all observable turbulent velocities on the $\sim 30$ kpc scale are below upper limits 
(with the exception of the re-simulation of E5A using bubbles, in which the turbulent velocity is of the order of the 2 lowest available upper limits within the same temperature range). 
This statistics is much more time-dependent
than the $\gamma$-ray fluxes, given the very small time scale associated with a
change of the average innermost temperature and small-scale gas velocity,
in response to AGN feedback events. In these runs, we verified that variations of factors $\sim 2-5$ in the average temperature,
and of factors $\sim 2-3$ in small-scale turbulence can easily be found
by comparing epochs separated by $\sim 100$ Myr. Radiative cooling and dissipation of turbulent motions on the one hand, and violent heating and driving of outflows from the AGN on the other can cause such negative and positive fluctuations of the measured temperatures and velocities inside $\sim 30 ~\rm kpc$. 
Taking this into account, we suggest that by using this technique it may be less
straightforward to unveil the presence of specific AGN modes in the real observation. Nevertheless, this approach appears still useful to limit the available energy budget of AGN feedback.

A complementary viable option in the near future is represented by X-ray spectroscopy
of Doppler-broadened iron lines in the ICM, such as the Fe XXIII line at $\sim 6.7 ~\rm keV$.
Several theoretical works already tested the impact of mergers
\citep[e.g.][]{2003AstL...29..783S,do05,2005ApJ...628..153B,va10tracers,2012MNRAS.422.2712Z}.
or AGN-feedback \citep[e.g.][] {2005ApJ...628..153B,heinz10} in shaping the Fe XXIII line with numerical simulations. 
In this case we can provide an overall view of the impact of various
AGN modes on the shape of the Fe XXIII line in our simulated
clusters.

\begin{figure*}
\includegraphics[width=0.33\textwidth]{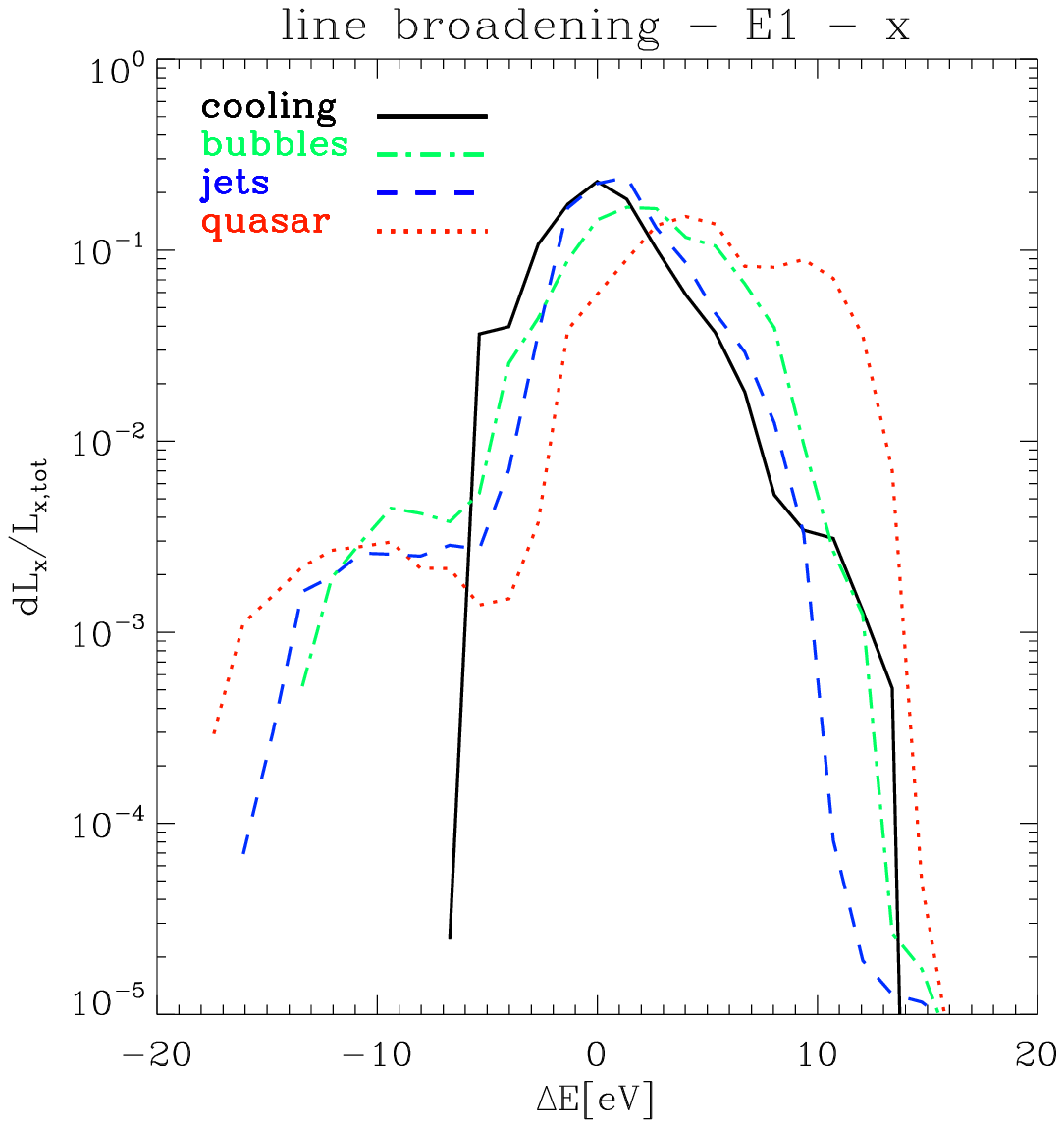}
\includegraphics[width=0.33\textwidth]{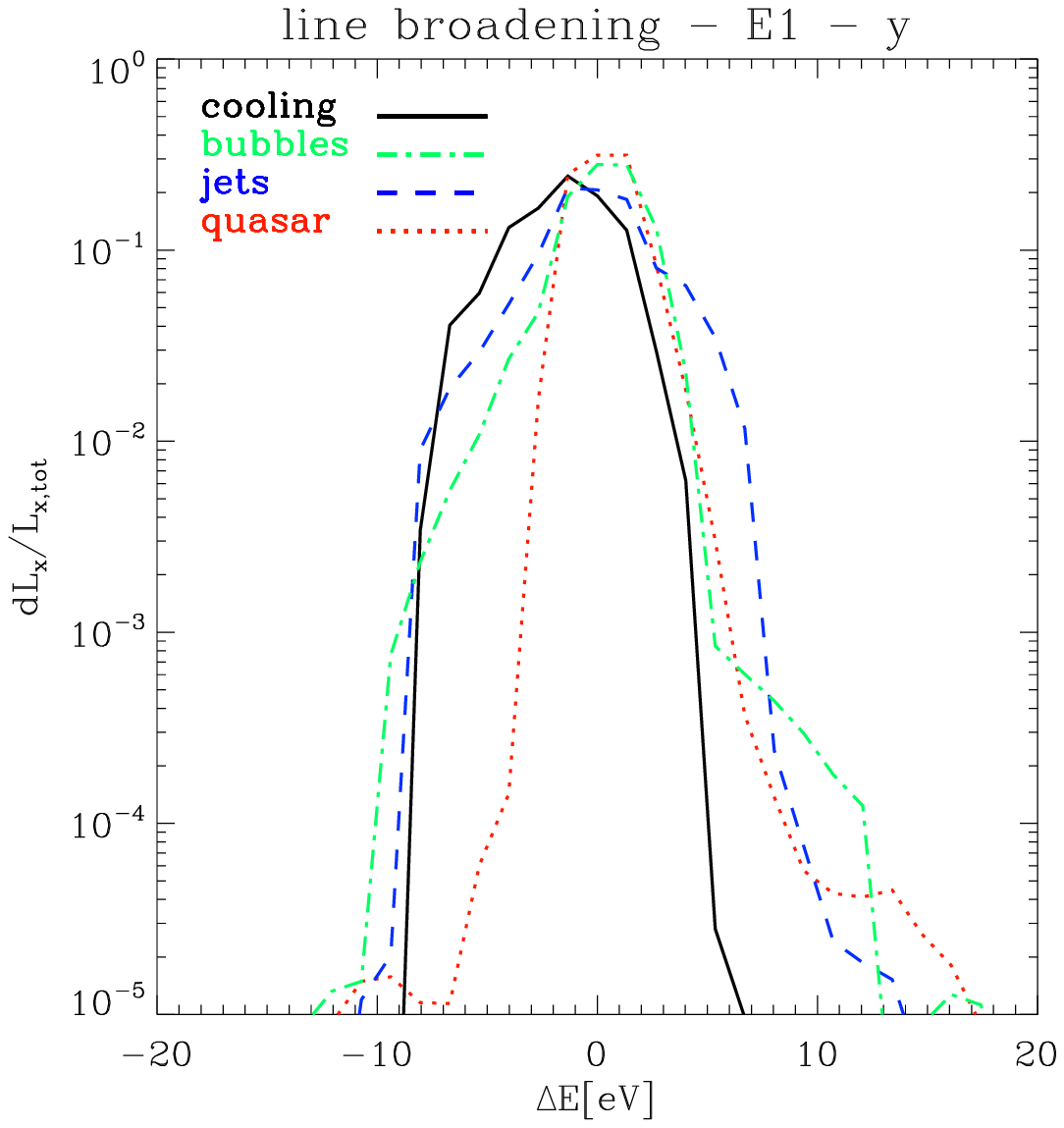}
\includegraphics[width=0.33\textwidth]{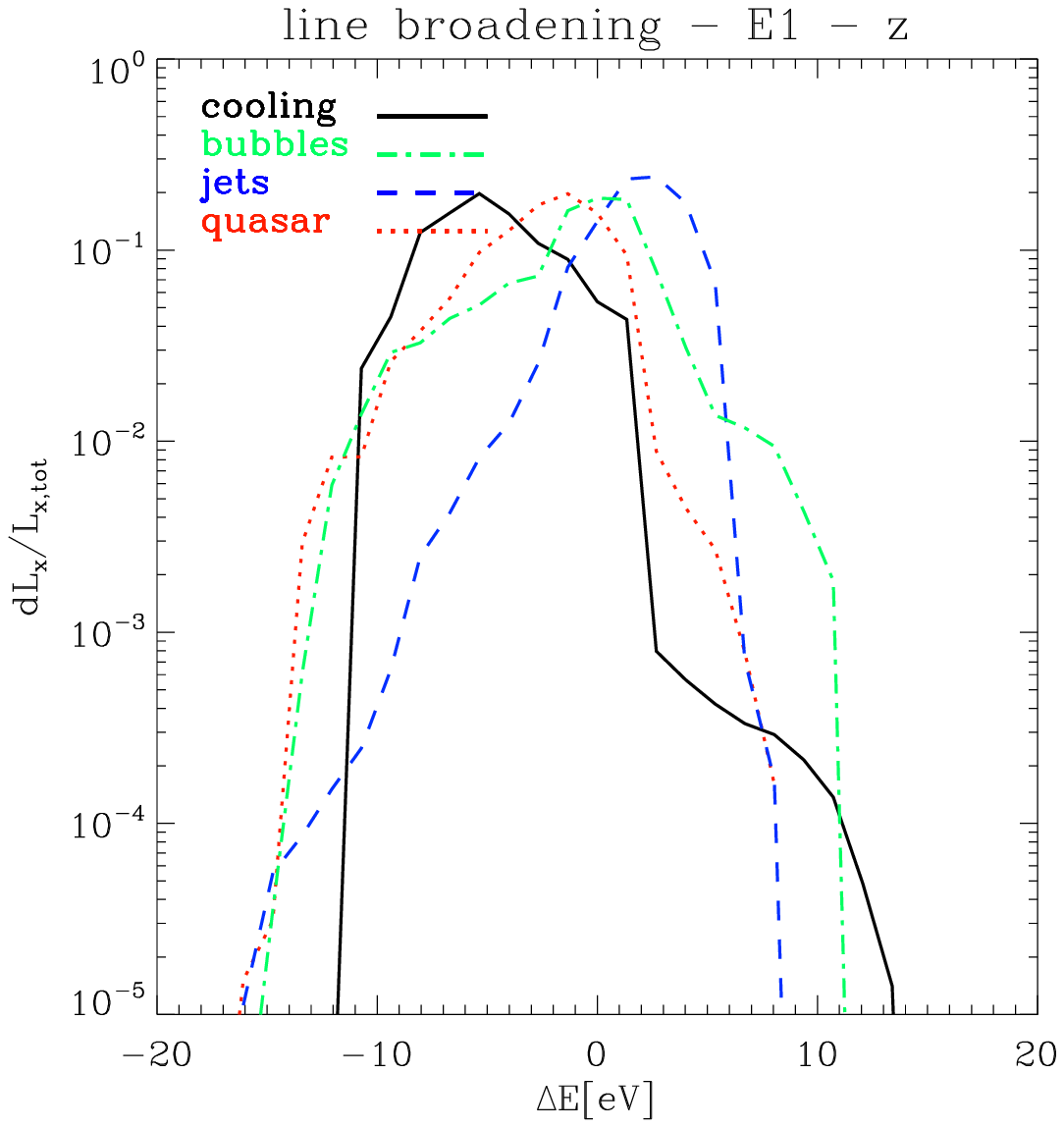}
\includegraphics[width=0.33\textwidth]{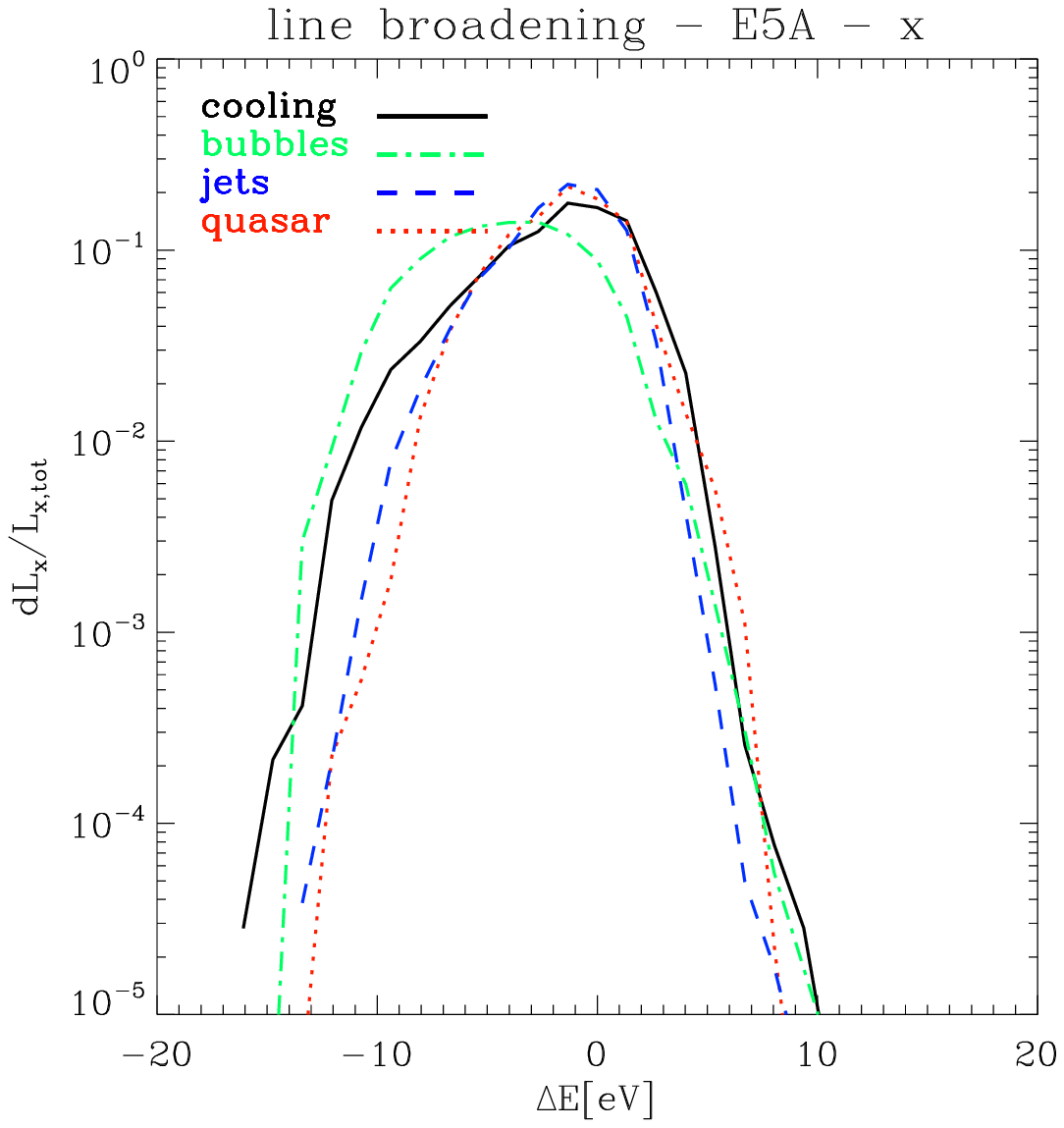}
\includegraphics[width=0.33\textwidth]{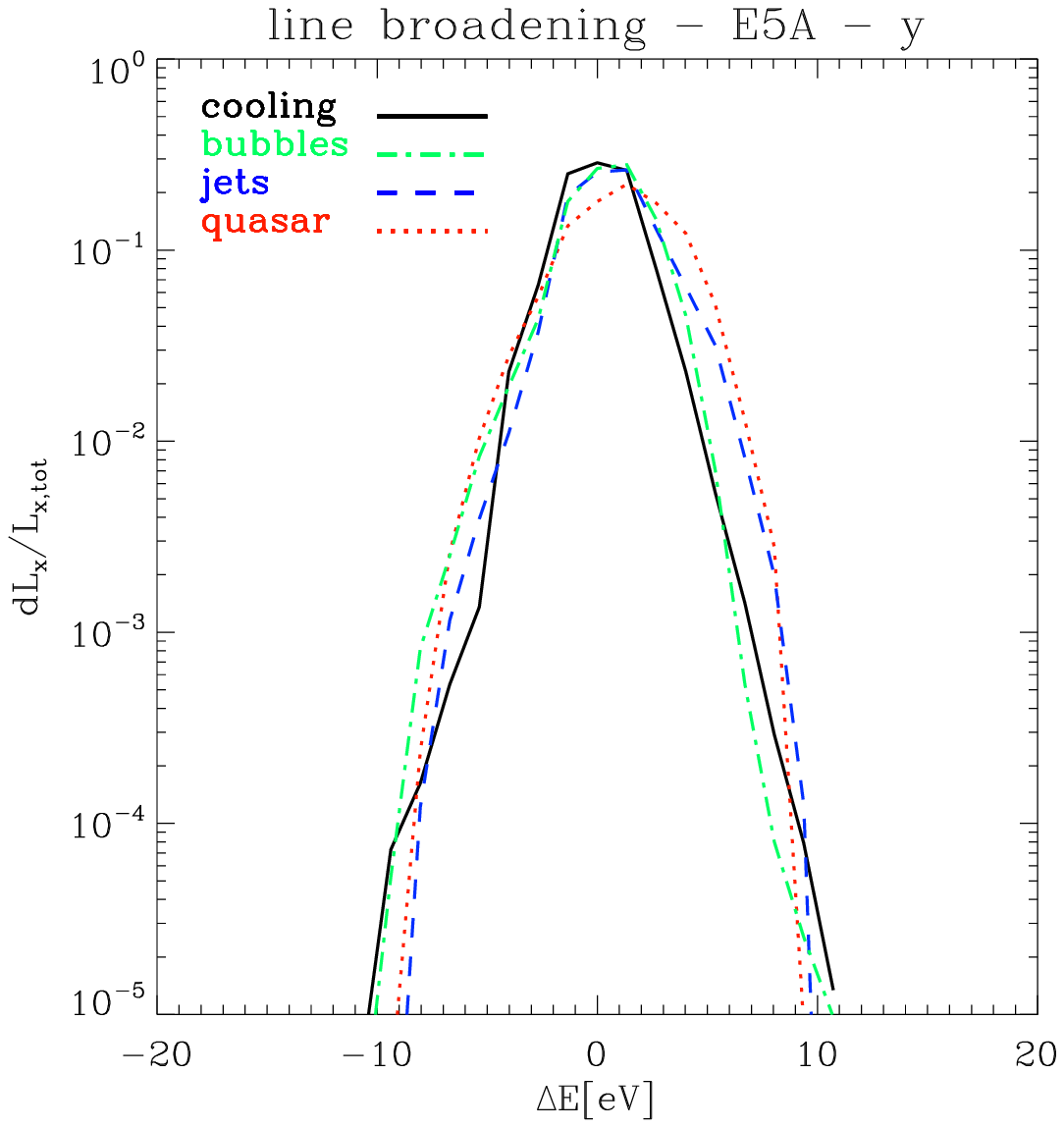}
\includegraphics[width=0.33\textwidth]{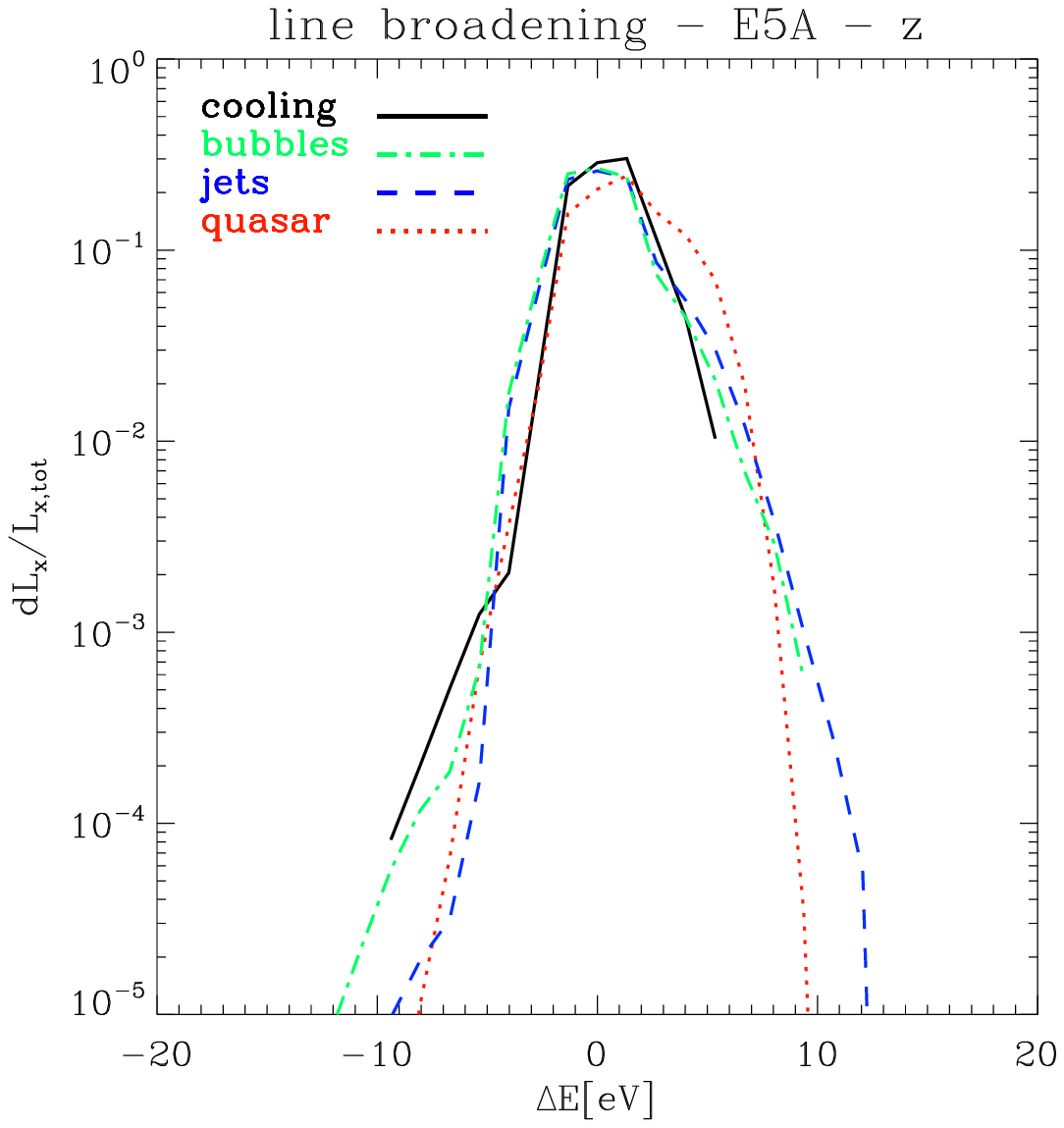}
\caption{Simulated Doppler broadening of FeXXIII line along three projection of cluster E1 (top) and E5A (bottom) at $z=0$. Both thermal and velocity broadening and considered. The emission is extracted from a projected area of $0.1 \times 0.1 ~\rm Mpc/h^2$. In each model the emission is normalized to the total inside the area.}
\label{fig:doppler}
\end{figure*}

In Fig.\ref{fig:doppler} we show the simulated broadening of the iron line from a region of $100 \times 100 ~\rm kpc^2$ centred on clusters
E1 and E5A, considering the three lines of sights. We also included
the effect of thermal broadening since feedback can simultaneously change the temperature of emitting gas along the line of sight.
To normalize the effect of very bright emission in the cooling run, we normalize all lines to their integrated luminosity within the field
of view.
The general effect of AGN feedback is to broaden the emission line, compared to the pure-cooling run, producing additional structures of high velocity tails (even if with very low associated luminosity)
with a FWHM $\sim 20-30 ~\rm eV$.
The different feedback modes present the largest difference along the z-direction (which is the direction of the jets). Run E1 presents the largest difference, since jets were active more recently in this system. 
However, we find that in general disentangling
the role of gravity-induced motions and of non-gravitational
processes is a very difficult task. Another problem is that the
high-velocity material must entrain enough high-emissivity material
to be detectable \citep[][]{2005ApJ...628..153B,heinz10}.
In order to efficiently deal with  "width-driven" 
or "separation-driven" features of the broadened emission, and to
disentangle the different sources of them,  sophisticated
techniques are being developed \citep[e.g.][and references therein]{2012arXiv1204.6058S}.

\section{Discussion}
\label{sec:discussion}

In this work we simulated the evolution of thermal and non-thermal properties
of a small sample of galaxy clusters  with a customized version
of the AMR code {\small ENZO} \citep[][]{no07,co11}.
The basic features of our two-fluid model have been presented
in \citet{scienzo}, while the implementation of shock-re-acceleration of CRs,
Coulomb and secondary losses and different modes of AGN feedback have
been introduced in this work (Sec.\ref{subsec:cr}-\ref{subsec:agn}).
To our knowledge, this is the first time that non-thermal
effects connected to cosmological shock waves and AGN feedback are studied
with cosmological simulations. We tested the performance
of a subset of recipes for AGN feedback (see, however, the Appendix
for a larger survey of models) against available observations
of thermal profiles derived from {\small CHANDRA} X-ray observations (Sec.\ref{subsec:xray}),
upper-limits of secondary $\gamma$-emission from {\small FERMI} observations
(Sec.\ref{subsec:gamma}) and upper-limits of turbulent motions from {\small XMM-Newton} observations (Sec.\ref{subsec:turbo}).
Our main result is that non-thermal observables allow us to assess the
reliability of each feedback mode against available upper-limits, and is thus complementary to X-ray observations. For a more
detailed discussion of the possible consequences of this method we refer
the reader to Sec.\ref{sec:conclusion}.

Here we list the main limitations of our current numerical method, and the way
in which our results can be affected or biased.

Firstly, our recipe to identify AGN-like cells in the simulated cluster volume (Sec.\ref{subsec:agn}) relies on the heuristic choice of a threshold in gas density ($n_{\rm min}=10^{-2} \rm cm^{-3}$). This choice is based on a comparison with the typical gas density of environments hosting super-massive BHs, reported in numerical work by other groups \citep[e.g.][]{2007MNRAS.380..877S,teyssier11}. The lack of the more consistent use of sink-particles to model at run-time the growth and accretion rate of BHs can be addressed with future developments in {\small ENZO 2.0} \citep[][]{co11}.

Secondly, we re-simulated a few clusters with several 
implementations of AGN feedback in order to compute the final budget of CR energy and their related $\gamma$-flux. These simulations are fairly expensive, and an extended survey of a large sample of clusters re-simulated with all variations
of AGN feedback is currently beyond our means. For this reason, an investigation of the presence of a bimodality in the distribution of
cool-core and non-cool-core clusters in our dataset was not possible. This 
is left to future work.

The recipe for radiative cooling is fairly idealized, and relies on the public implementation of equilibrium cooling in a fully-ionized H-He plasma with
constant metallicity. Even though a more self-consistent inclusion of metal
ejection from supernovae and galactic winds and line cooling can exacerbate
the cooling catastrophe in radiative simulations \citep[][]{dubois11}, the overall 
energy budget required from AGN feedback in our simulations (and its influence
on the CR energy budget) should not be significantly affected. 
However, the inclusion of metals in our
 description of the ICM can provide an additional way of 
 studying turbulent motions in the ICM, since successful implementations
 of feedback must yield the bimodality of metallicity
 profiles observed in cool-core and non-cool-core clusters \citep[e.g.][]{2008A&A...487..461L}. 

The neglect of star formation may artificially increase (by a $\sim 20-30$ per cent, \citealt[e.g.][and references therein]{2011ApJ...731....6S,2011ApJ...731...11C}) the mass of the hot gas phase in our runs here. At the same
time, the inclusion of feedback from star formation and supernovae might somewhat reduce the energy budget
required from AGN \citep[e.g.][]{2003MNRAS.342.1025T,2003MNRAS.339.1117V,2007ApJ...668....1N,short12}. 
We will investigate this topic in the near future, using simulations with the effect of star formation
included.

The inclusion of magnetic fields in these simulations is not expected to change
the overall thermal structure of the ICM \citep{do99,xu09,co11}. It can, however, affect the mixing of jets and bubbles with the ICM \citep[][]{oj10,2012ApJ...750..166M}.

A final important caveat concerns the spatial resolution of our runs. The minimum cell size of $25 ~\rm kpc/h$ (almost uniform within the AMR region) is sufficient to capture shock waves and turbulent features related to cluster mergers \citep[as in][]{va11turbo}. However, it is barely enough to capture the interplay between the AGN region and the cluster core. Modelling the accretion rate on the central galaxy at the centre of the cooling radius \citep[][]{2012ApJ...747...26L}, the small-scale interaction between jets \citep[e.g.][]{2006MNRAS.373L..65H,2010MNRAS.407.1277M,gaspari12} or inflated bubbles \citep[e.g.][]{2008ApJ...686..927S,2008MNRAS.387.1403S}, requires 
a resolution of the order of $< \rm kpc$.
Moreover, to resolve the turbulence excited by cluster mergers, sloshing and AGN-jets in the same simulation, one would need to cover scales ranging from $\rm R_{\rm vir} \sim 3 ~\rm Mpc$ down to the presumed scale of physical dissipation at $\sim 0.1$ kpc, in a rather uniform way, for a range of scales of $10^3-10^4$.
 
\section{Conclusions}
\label{sec:conclusion}

What can be learned from this exercise of comparing
thermal and non-thermal observables of simulated and observed clusters? 
We argue that the approach outlined here can put robust
constraints on the energetics, duty cycle, mechanism and epochs of
feedback from AGN.
Even if with present data it is not yet possible to reject specific 
implementations of feedback, it can be a powerful approach in the near future.
The amount of CR injection after each AGN burst is assumed to be given by the
shock acceleration efficiency of diffusive shock acceleration in \citep{kj07,2012arXiv1206.1360C}.
The other mechanisms of dynamical (e.g. advection with the fluid) or energy (e.g. secondary losses) evolution of CRs depend the dynamics of the simulated ICM. All additional mechanisms of CR injection that we neglected here (e.g. direct injection from supernovae, galactic winds, magnetic reconnection) cannot but {\it increase} the budget
of CRs in the ICM, even if not substantially. Therefore, it is likely that our limits of $\gamma$-emission from CRs are slightly low.
We also remark that other mechanisms of CRs, such has CRs diffusion \citep[e.g.][]{hl00,ju08},
 should be negligible for the $>25 ~\rm kpc/h$ scales of interest here \citep[e.g.][and references therein]{blasi07}{\footnote {In \citet{2011A&A...527A..99E}  it is argued that if CR can stream along a quasi radial magnetic field much faster than the Alfv\'{e}n speed, the CR-energy density in radio-quiet clusters would be greatly reduced. However, it seems that such an hypothesis can be excluded on theoretical grounds \citep[][]{1981A&A....98..161A,1989ApJ...336..243S,1994ApJS...90..929S,2004ApJ...604..671F}, and also  based on a large number of Faraday Rotation \citep[][]{ev03,mu04,2004JKAS...37..337C,gu08,bo10,vacca10} and polarization \citep[][]{bonafede11}  data from nearby clusters, that seem to exclude a systematic difference in the topology of the ICM magnetic fields between clusters with and without large-scale emission. }}.

Therefore, in the near future a set of $\gamma$-observation significantly below these numerical estimates for jets/AGN feedback may imply one of the following possibilities:
\begin{itemize}
\item the real power per event for each AGN burst is substantially lower than what we assumed here ($\sim 10^{44}-10^{45} \rm erg/s$) with a different duty cycle. However, such large powers have been inferred from observations
in several cases \citep[e.g.][and references therein]{2007ARA&A..45..117M,2012MNRAS.422.2213S}.

\item the acceleration efficiency
of CRs at $M<10$ shocks (which are typical in AGN bursts and cluster mergers) is lower than what we assumed here. Indeed, 
the details of
particle acceleration for $2 \leq M \leq 10$
are not yet robustly constrained by theory due to the difficulty of modelling the large range of
spatial and temporal scales involved in the diffusive acceleration at such shocks \citep[e.g.][]{kr10}. More recently, several groups employing particle-in-cells methods investigated
additional acceleration mechanisms for particles at shocks (e.g. shock drift acceleration),
suggesting the possibilities of a different dependence on Mach number \citep{ga11}.
In addition, there is now growing evidence that the non-linear diffusive shock acceleration model has
to be slightly revised to fully explain recent data from supernovae \citep[][]{2012arXiv1206.1360C}. 

\item the actual quenching of catastrophic cooling in real clusters can happen over time through a {\it mixture} of "violent" (jets, quasar) and more "quiet" (e.g. bubbles) episodes. A physically motivated mixture of these different violent and quiet phases may in principle quench the cooling catastrophe without injecting too large CR-energy in the ICM. Theoretical work, indeed, suggests that this is a viable
possibility \citep[e.g.][]{2007MNRAS.380..877S,mcc2008,short12}. 
\end{itemize}

Based on the results presented in this paper, we argue that in the future a careful modelling of non-thermal observables will be important to improve our understanding of AGN feedback. Despite the energy that is theoretically
available for AGN feedback, the AGN cannot be arbitrarily impulsive or continuous because in both cases
this will affect observable non-thermal phenomena
(such as turbulence and the injection of CR-energy).
In particular, given the extremely long time for Coulomb and hadronic losses of CR-protons in the ICM, the investigation of (lack of) $\gamma$-emission or radio-emission from secondary particles is a powerful tool to unravel
feedback during the earliest phases of structure formation.

\section*{acknowledgements}

We thank the referee for the very helpful comments, which
improved the final quality of the paper.
F.V. and M.B. acknowledge support
from the grant FOR1254 from the Deutsche Forschungsgemeinschaft. 
F.V. acknowledges the 
usage of computational resources under the CINECA-INAF 2008-2010 agreement, and at the at the John-Neumann Institut at the
Forschungszentrum J\"{u}lich. 
We thank J. Donnert, D. Collins, M. Gaspari, F. Brighenti \& K. Dolag and G. Brunetti for very useful scientific discussions, and J. Sanders for
kindly providing the observed data-points
of Fig.\ref{fig:sanders_limits}.
We gratefully acknowledge the {\it ENZO} development group for providing extremely helpful and well-maintained on-line documentation and tutorials (http://lca.ucsd.edu/software/enzo/).

\bibliographystyle{mnras}
\bibliography{franco}

\bigskip

\appendix

\section{Tests with varying parameters}

In this Appendix we show results of a more extended  study of the variation of thermal and CR properties of one cluster run (H10), 
with a more systematic exploration of the parameter space of our implementation of AGN feedback. 
To save computational time, all tests
in this case have been performed using the standard mesh refinement
strategy, where the grid is refined in regions where the gas/DM over-density 
is larger
than 3 times the overdensity of the parent cell (the differences in the thermal properties inside $R_{\rm 200}$ are 
very small compared to the AMR criteria used in the main article, while
in general the properties of accretion shocks and turbulent motions tend to 
be significantly affected by resolution issues and numerical dissipation,
 as discussed in \citet{va09turbo} and \citet{va11entropy}).

\begin{figure*}
\includegraphics[width=0.95\textwidth]{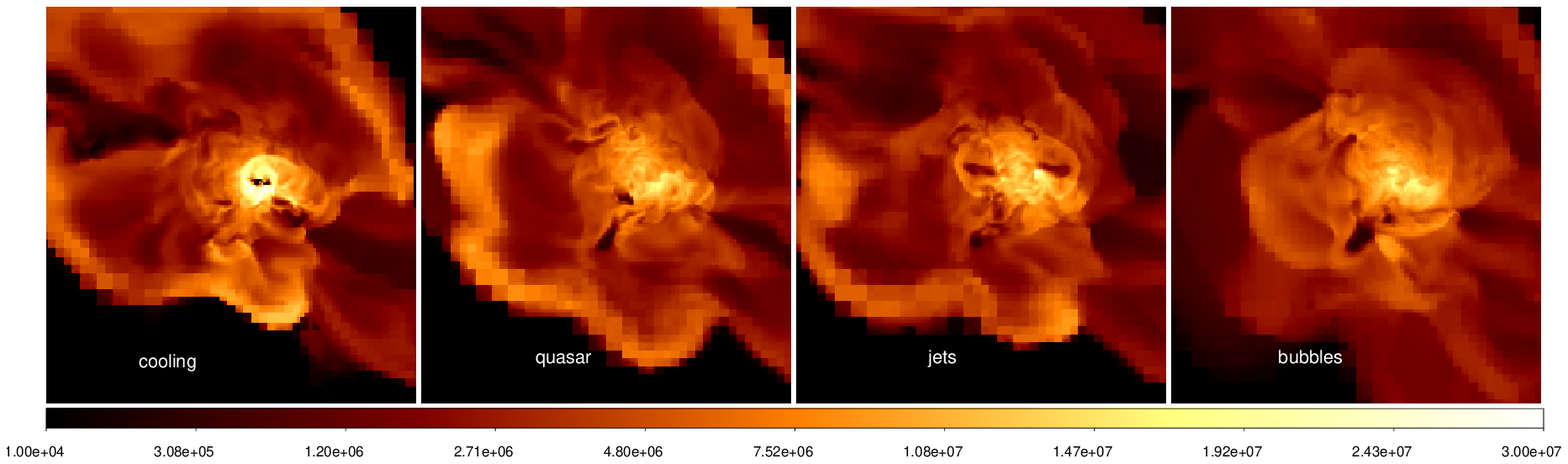}
\includegraphics[width=0.95\textwidth]{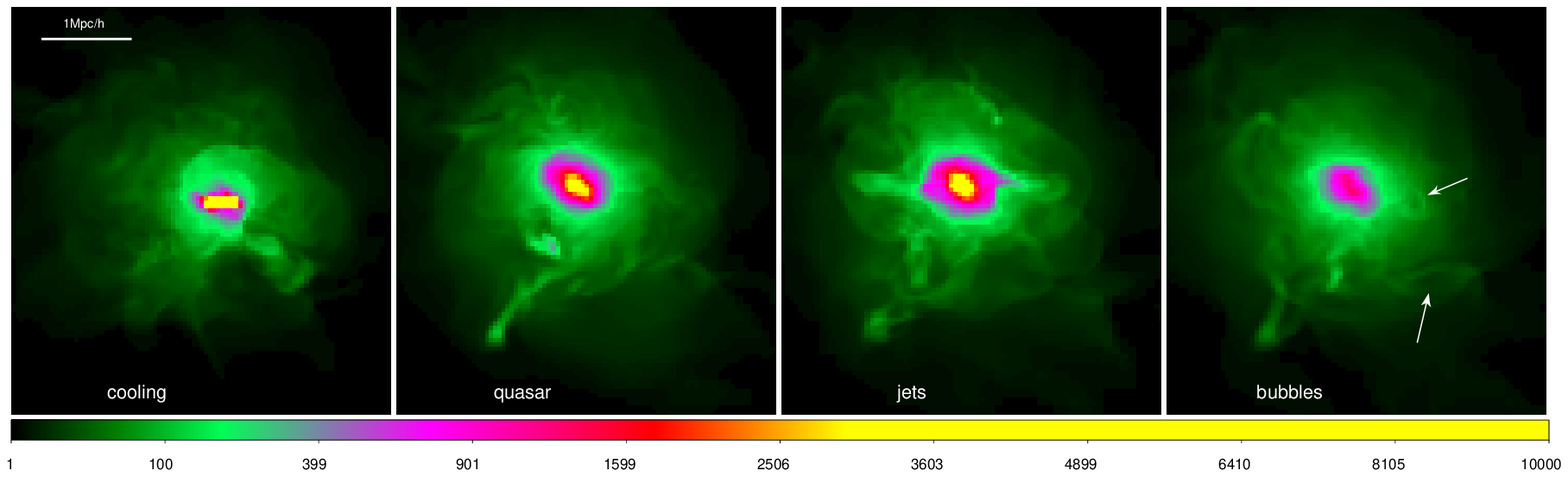}
\caption{Top panels: slice of temperature (for a slab of $25 ~\rm kpc/h$ through the cluster centre) for cluster H10 at $z=0$, re-simulated with 4 different physical models. Each image has a side $\approx 5  ~\rm Mpc/h$. Bottom panels: slice of gas density for the same re-simulations of cluster H10 (note that the 
size of the images has been reduced ($\approx 2.5  ~\rm Mpc/h$), to highlight the features in the cluster core). The arrows show the location of buoyant bubbles in the {\it B2} mode.}
\label{fig:appendix1}
\end{figure*}

Figure \ref{fig:appendix1} shows the projected gas density and a temperature cut for four re-simulations of this cluster for the pure cooling case ({\it C}) and for 3 feedback models with similar feedback energy budget ({\it A2}, {\it K2} and {\it B2}). Despite the fact that the outer accretion shocks are poorly described because of the AMR strategy employed here, the innermost region where the AGN feedback
plays a role are resolved similarly to the runs studied in the main article. 

\begin{figure*}
\includegraphics[width=0.95\textwidth,height=0.4\textheight]{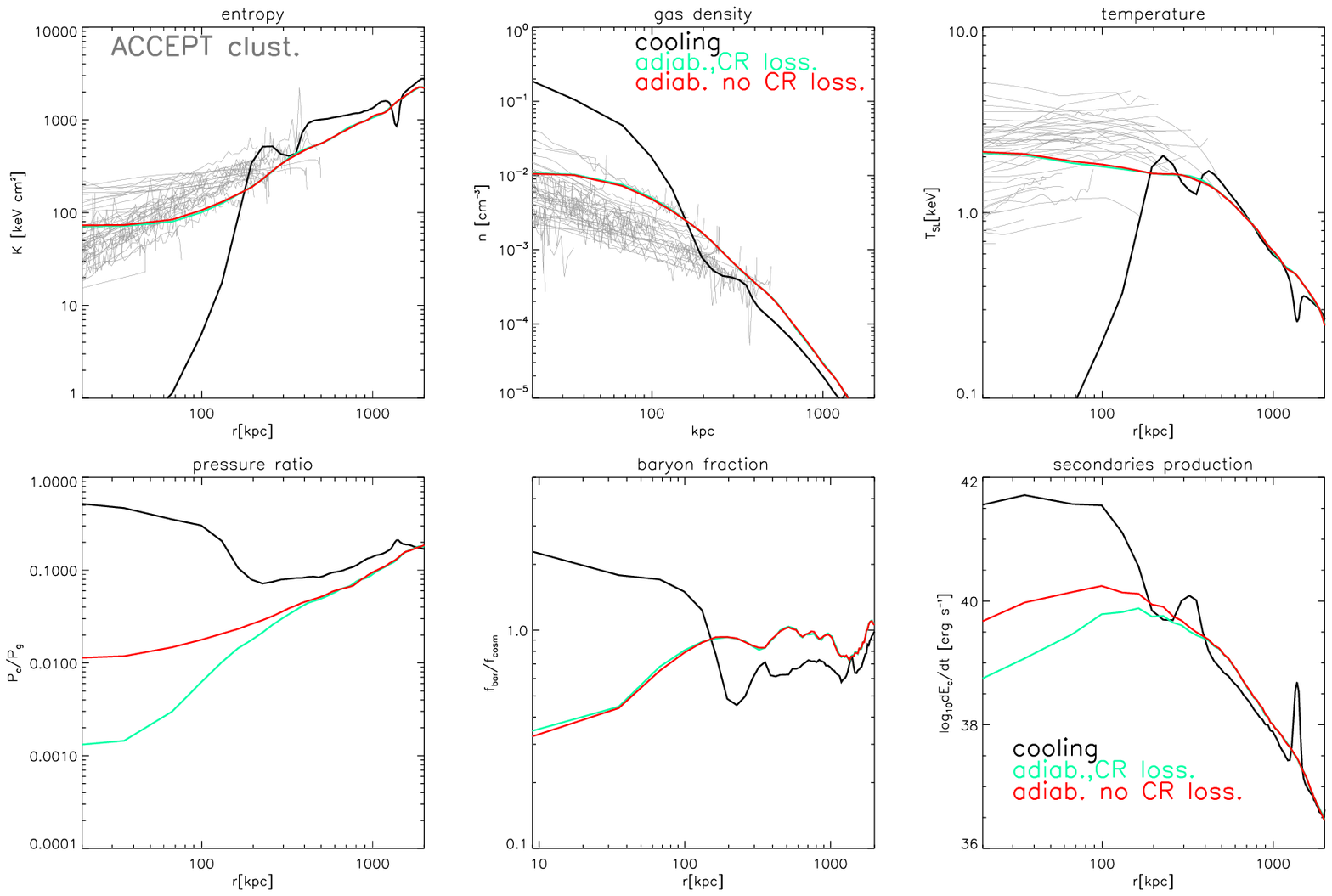}
\caption{Radial profiles of gas entropy, gas density, gas temperature, pressure ratio $P_{\rm cr}/P_{\rm g}$, baryon fraction and secondary injection for the re-simulations of cluster H10 employing radiative cooling only, and non-radiative physics with and without CR losses.
We additionally show as grey line the profile of the clusters in the sample of \citep{cav09}.} 
\label{fig:appendix2}
\end{figure*}

\begin{figure*}
\includegraphics[width=0.95\textwidth,height=0.4\textheight]{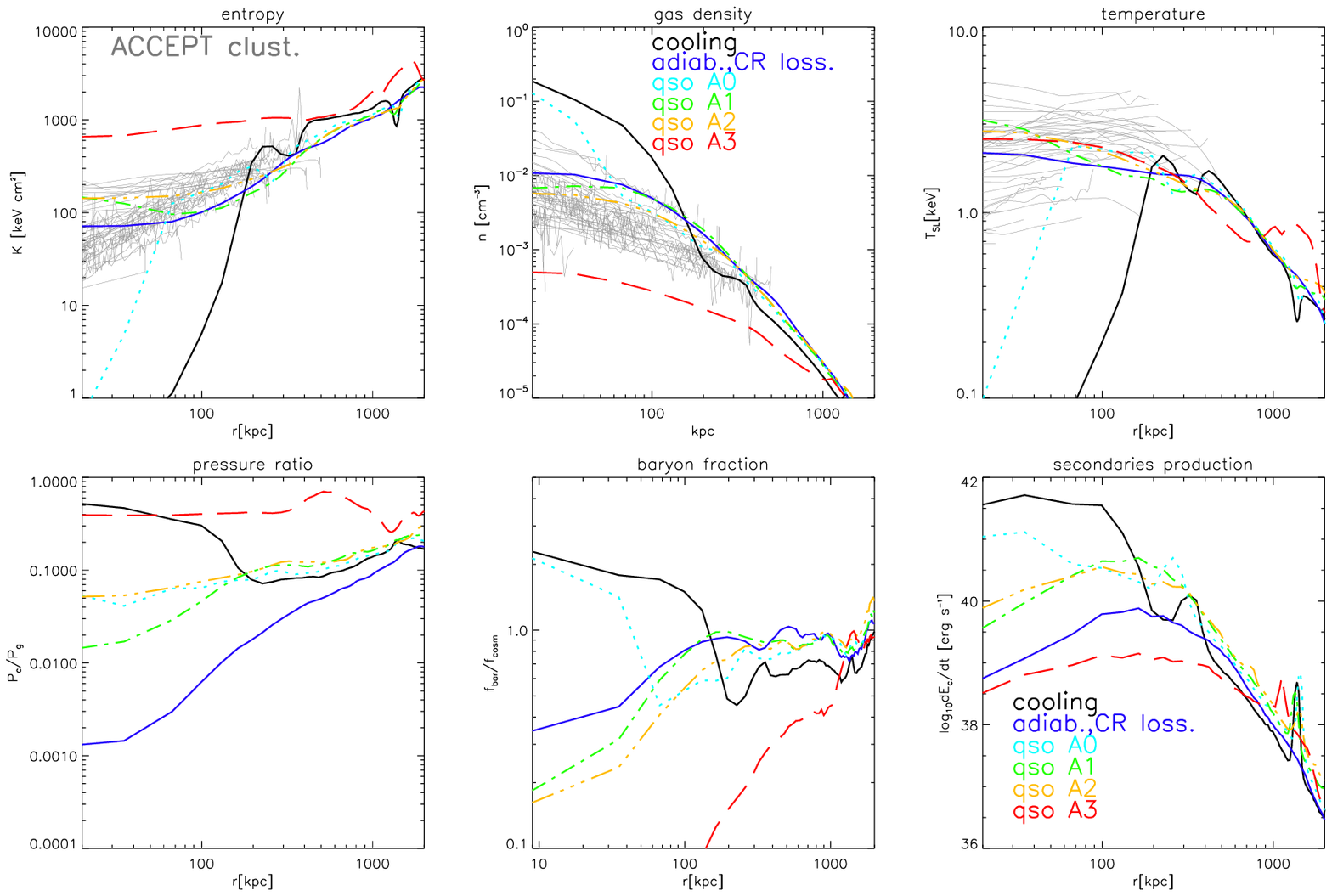}
\caption{Same as in Fig.\ref{fig:appendix2}, but for re-simulations
of H10 with different implementations of thermal feedback from quasars.}
\label{fig:appendix3}
\end{figure*}

\begin{figure*}
\includegraphics[width=0.95\textwidth,height=0.4\textheight]{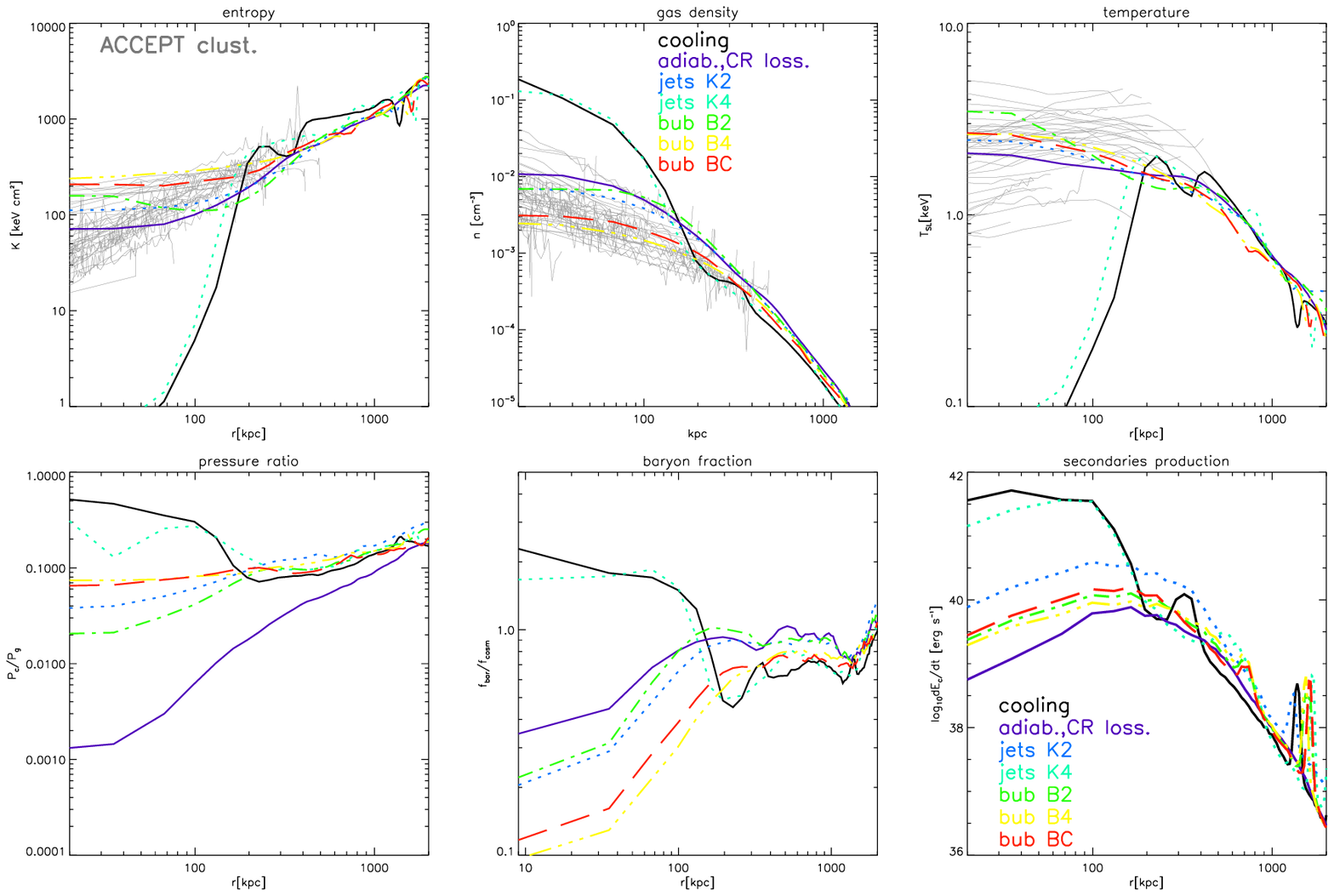}
\caption{Same as in Fig.\ref{fig:appendix2}, but for re-simulations
of H10 with different implementations of jet or bubble feedback.}
\label{fig:appendix4}
\end{figure*}

Table A1 lists all re-simulations of H10 performed, with details on the adopted feedback model. 
In the ensemble of profiles shown in Fig.\ref{fig:appendix2}-\ref{fig:appendix4} we show all profiles of thermal and CRs related properties for all re-simulations.

We compared one standard non-radiative run with CR feedback only from cosmological shock acceleration (as in \citealt{scienzo}, run  {\it NR0}), one run where also the Coulomb and secondary losses are modelled ({\it NR1}) and the pure cooling case with losses for CRs ({\it C}).

Some results can be derived from this preliminary comparison of feedback modes, in additon to those already discussed in the main article. a) The adoption of a more standard AMR criterion still gives internal profiles and trends with AGN feedback very similar to the more expensive resimulations with AMR based also on velocity jumps. b) The thermal properties of the cluster are unchanged  by the adoption of CR losses at run time (Fig A2). The amount of CR-energy within the innermost $100 ~\rm kpc$ and the production rate of secondary particles (Eq.\ref{eq:secondary}) are however reduced by a factor $\sim 10$ due to CR losses. However, the inclusion of CR-losses is expected to reduce the cooling rate in radiative simulations, owing to the larger gas density attained there. However, CR-losses are not able to stem the cooling flow, unless other heating mechanisms by CRs \citep[e.g. heating from Alfv\'{e}n waves excited by the streaming of CRs in a $\sim 1-10 \mu G$ magnetic field][]{fuj11} are at work. c) In the case of quasar feedback, an energy per event of $10^{60} \rm erg$ (i.e. 10 times larger than the fiducial model for H10 discussed in the main paper) in a cluster
of this mass creates a baryon poor system, and a unrealistically
large ratio of CR to gas energy ($\sim 40$ percent within the virial radius). d) For the quasar mode (Fig. A3), an energy per event of $10^{58} \rm erg$
(i.e. 10 times smaller than the fiducial model for H10 discussed in the main paper) can produce an acceptable match to the observed X-ray profiles, similar to what we reported for the two largest cluster masses (Sec. \ref{subsec:xray}). If AGN feedback is switched off too early ($z=1$), a cooling catastrophe develops
only for a small region, $\leq 50 ~\rm kpc$. e) In the jet mode (Fig.A4), too slow jets ($\leq 10^{2} ~\rm km/s$) are unable to quench the
cooling flow. f) At this rather poor resolution, the investigated differences in the initial density contrast of bubbles (Fig.A4), or in their initial energy composition (i.e. thermal energy or CR-energy dominated) produce extremely similar result on the thermal and CR properties of the cluster at $z=0$.

\begin{table}
\label{tab:tab_app}
\caption{List of the physical models adopted in our runs. Column 1: identification name.
C2: cooling. C3: details of feedback model. C4: mnemonics.}
\centering \tabcolsep 5pt 
\begin{tabular}{c|c|c|c}
  ID & cooling & feedback mode & mnemonic\\  \hline
  NR0 & no  & no & no cooling,no.CR losses \\
  NR1 & no  & no & no cooling.,CR losses \\ 
  C  & yes &  no & cooling\\ \hline
  A1 & yes &  thermal, $T_{\rm AGN}=10^{7} \rm K$ & quasar low\\
  A2 & yes &  thermal, $T_{\rm AGN}=10^{8} \rm K$  & quasar med.\\
  A3 & yes &  thermal, $T_{\rm AGN}=10^{9} \rm K$  & quasar high.\\
  A0 & yes &  thermal, $T_{\rm AGN}=10^{8} \rm K$ $z \geq 1$ & early quasar \\
  K2 & yes &  kinetic, $v_{\rm jet}=10^{3} ~\rm km/s$  & jets high\\
  K4 & yes &  kinetic, $v_{\rm jet}=10^{2} ~\rm km/s$  & jets low\\
  B2 & yes & buoyant, $\delta_{\rm bbl}=0.1$ & bubbles \\
  B4 & yes & buoyant, $\delta_{\rm bbl}=0.05$ & bubbles \\ 	
  BC & yes & buoyant, $\delta_{\rm bbl}=0.05$, $\phi_{\rm cr}=10$ & bubbles \\ 	
\end{tabular}
\end{table}

 \end{document}